\documentclass[prc,superscriptaddress,twocolumn]{revtex4}

\usepackage{hyperref}
\usepackage{amssymb}
\usepackage{graphicx}
\usepackage{mathrsfs}
\usepackage{hyperref}
\usepackage[usenames,dvipsnames,svgnames]{xcolor}

\begin{document}

\title{Effects of bulk viscosity and hadronic rescattering
in heavy ion collisions at RHIC and LHC}
\author{Sangwook Ryu}
\affiliation{Department of Physics, McGill University,
3600 rue University, Montreal, Quebec H3A 2T8, Canada}
\affiliation{Frankfurt Institute for Advanced Studies,
Ruth-Moufang 1, 60438 Frankfurt, Germany}
\author{Jean-Fran\c{c}ois Paquet}
\affiliation{Department of Physics \& Astronomy, Stony Brook University,
Stony Brook, NY 11794, USA}
\affiliation{Department of Physics, McGill University,
3600 rue University, Montreal, Quebec H3A 2T8, Canada}
\author{Chun Shen}
\affiliation{Department of Physics, McGill University,
3600 rue University, Montreal, Quebec H3A 2T8, Canada}
\author{\mbox{Gabriel Denicol}}
\affiliation{Instituto de F\'{\i}sica, Universidade Federal Fluminense,
UFF, Niter\'{o}i, 24210-346, RJ, Brazil}
\author{Bj\"orn Schenke}
\affiliation{Physics Department, Brookhaven National Laboratory,
Upton, NY 11973, USA}
\author{Sangyong Jeon}
\affiliation{Department of Physics, McGill University,
3600 rue University, Montreal, Quebec H3A 2T8, Canada}
\author{Charles Gale}
\affiliation{Department of Physics, McGill University,
3600 rue University, Montreal, Quebec H3A 2T8, Canada}
\date{\today}

\begin{abstract}
We describe ultra-relativistic heavy ion collisions at RHIC and the LHC
with a hybrid model using the IP-Glasma model for the earliest stage
and viscous hydrodynamics and microscopic transport
for the later stages of the collision.
We demonstrate that within this framework the bulk viscosity of the plasma
plays an important role in describing the experimentally observed radial flow
and azimuthal anisotropy simultaneously.
We further investigate the dependence of observables on the temperature
below which we employ the microscopic transport description.
\end{abstract}
\maketitle
\section{Introduction}

Ultra-relativistic heavy ion collisions carried out
at the Relativistic Heavy-ion Collider (RHIC)
and the large Hadron Collider (LHC) are unequaled tools
to study the many-body properties of quantum chromodynamics (QCD),
in particular its high-temperature deconfined phase
known as the quark-gluon plasma (QGP)~\cite{gale:2013,deSouza:2015ena}.
Since the QGP is only produced for a very short
time and cannot be observed directly,
extracting its properties from heavy ion measurements is a major challenge
that requires modelling the many stages of the collision:
the pre-equilibrium dynamics of the system,
the rapid expansion and cooling of the QGP
and the dynamics of the dilute hadronic matter
that is eventually measured by the experiments. 

In the past decade,
hydrodynamic models have been applied
with great success to describe the distribution of soft hadrons produced in
heavy ion collisions at RHIC and the LHC~\cite{gale:2013,deSouza:2015ena}.
The foremost experimental discovery made
using these models was that the QGP displays remarkable transport properties,
with one of the smallest shear viscosity to entropy density ratio ever observed
\cite{Gyulassy:2004zy,whitepaper1,whitepaper2,whitepaper3,whitepaper4}.
A more precise determination of the transport properties of QCD matter,
including their non-trivial temperature dependence,
is one of the primary goals of the heavy ion research program.

For a long time, shear viscosity was considered to be
the dominant source of dissipation for the QGP
produced in heavy ion collisions
\cite{Romatschke:2007mq,MUSIC,Song:2010aq,Song:2011qa,Arnold:2006bv}.
Nevertheless, there are theoretical indications
that bulk viscosity can become large around the QCD crossover region
\cite{Meyer:2007dy,Karsch:2007jc,Buchel:2007mf,
Kharzeev:2008tra,NoronhaHostler:2008ju}
and can significantly affect the evolution of the QGP
\cite{Torrieri:2008ip,Rajagopal:2009yw,
Denicol:2009am,Habich:2014tpa,Denicol:2015bpa}.
Early investigations on the effect of bulk viscosity
using realistic hydrodynamic simulations often assumed small values
for this transport coefficient \cite{Bozek:2009dw,Bozek:2012qs,Song:2009rh}
and found modest effects.
Other studies focused on the effects of dissipative corrections
due to bulk viscosity in the particlization \cite{Huovinen:2012is} of
the hadron resonance gas \cite{Monnai:2009ad,Dusling:2011fd,
Noronha-Hostler:2013gga,Noronha-Hostler:2014dqa}.
Whether a large bulk viscosity can be reconciled
with the current theoretical description of heavy ion collisions
is a topic of great interest to the field.

Recent calculations done in Ref.~\cite{Ryu:2015vwa} have addressed this
issue within a modern hydrodynamic description,
finding that a large bulk viscosity around the phase transition region is
essential to describe simultaneously the multiplicity
and average transverse momentum of charged hadrons.
This finding was made
using IP-Glasma initial conditions \cite{Schenke:2012wb},
second order hydrodynamic equations \cite{MUSIC,Denicol:2014vaa},
and a transport description
of the late stages of the collision~\cite{UrQMD1,Bleicher:1999xi}.
Similar conclusions about the importance of bulk viscosity
have since been reached by calculations
employing different initial state models that, similarly to IP-Glasma,
also exhibit large sub-nucleonic energy density fluctuations
\cite{Denicol:2015nhu,Bernhard:2016tnd}.

The goal of this paper is to expand on the results
presented in Ref.~\cite{Ryu:2015vwa},
offering a more detailed overview on the effect of bulk viscosity
on other heavy ion observables and different collision energies.
In particular, the effects of late stage hadronic rescattering
\cite{Bass:2000hy,Teaney:2001fl} will be discussed in greater details.
These have been investigated in
a number of previous publications
\cite{Song:2010aq,Song:2011qa,Song:2013qma,Zhu:2015dfa,
Hirano:2005xf,Hirano:2007ei,Hirano:2010jg,Takeuchi:2015ana,
Nonaka:2007hy,Petersen:2008in,Nonaka:2010hc,Petersen:2010tr},
where they were found to be especially important
to provide a reasonable description
of the hadronic chemistry of heavy ion collisions,
specially for heavier baryons.
Here, we shall perform a systematic study of the effects of the switching
temperature between the hydrodynamic simulation and the transport model,
showing that this parameter has a significant effect
on the momentum distribution of protons and multi-strange hadrons. 

The rest of this paper is organized as follows.
In Section~\ref{sec:model},
we explain each of the components of our model and
show how they are combined to give an integrated description.
In Section~\ref{sec:results},
we compare the results of our calculations with experimental data,
focusing on the effect of bulk viscosity and hadronic rescattering.
We show that a finite bulk viscosity resolves the tension between
the observed multiplicity and the average transverse momentum.
We also show that the hadronic cascade is an important ingredient
for the description of hadronic chemistry.
We summarize our results and discuss implications in Section~\ref{sec:summary}.

\section{Model}
\label{sec:model}

The theoretical framework used in the present paper
can be divided in four parts:
the pre-equilibrium dynamics described with the IP-Glasma model,
the hydrodynamical evolution, the transition from fluid to particles and
the final hadronic transport.

\subsection{Pre-equilibrium with IP-Glasma}

The IP-Glasma model \cite{Schenke:2012wb} describes
the pre-equilibrium dynamics of a large number of low-$x$ gluons
by the classical Yang-Mills equation.
The high-$x$ partons serve as the color sources
for the initial gluon fields before the collision.
For each nucleus,
color charges $\rho_i^a ( \textbf{x}'_T )$
with color index $a$ at a lattice site $i$ in
the $x^{\pm}$-direction are sampled
according to a Gaussian distribution, satisfying
\begin{equation}
	\langle \rho_i^a ( \textbf{x}'_T ) \rho_j^b ( \textbf{x}''_T ) \rangle
		= g^2 \mu_A^2 ( \textbf{x}'_T ) \,
		\delta^{ab} \, \frac{ \delta^{ij} }{ N_L } \,
		\delta^{(2)} ( \textbf{x}'_T - \textbf{x}''_T ) \, ,
\end{equation}
where $N_L$ is the number of lattice sites in the $x^{\pm}$-direction. 
The value $N_L= 100$ is used in this work.
The average color charge density per unit transverse area
$g^2 \mu_A^2 ( \textbf{x}_T )$
is proportional to the saturation scale $Q_{s,A}^2 ( \textbf{x}_T )$
determined in the impact parameter dependent saturation model
(IP-Sat)~\cite{Bartels:2002msat,Kowalski:2003ipsat}
\begin{eqnarray}
	Q_{s,A}^2 ( \textbf{x}_T ) & = & \frac{2 \pi^2}{N_c}
		\alpha_s (\mu^2 (r_s^2)) \,
		x f_g ( x, \mu^2 (r_s^2)) \nonumber\\
		& & \times \sum_{i=1}^A \frac{1}{2\pi \sigma_0^2}
		\exp{\left[ - \frac{(\textbf{x}_T - \textbf{x}_{T,i})^2}{2 \sigma_0^2}
		\right]} \nonumber \\
		&=& \frac{2}{r_s^2} \\
	\mu^2 ( r^2 ) & = & \frac{C}{r^2} + \mu_0^2
\end{eqnarray}
where $f_g (x,\mu^2)$ is the gluon distribution function in a nucleon
and $\mu_0$ is a momentum scale
at which the gluon distribution has a form of~\cite{Bartels:2002msat}
\begin{equation}
x f_g ( x, \mu^2 = \mu_0^2 ) = A_g x^{- \lambda_g} ( 1 - x )^{5.6}\;.
\end{equation}
The gluon distribution at an arbitrary momentum scale $\mu>\mu_0$ is obtained
from DGLAP evolution.
Each nucleon is assumed to have a Gaussian shape
with width $\sigma_0$ in the transverse plane.
The parameters are determined
to fit HERA deep inelastic scattering data \cite{Kowalski:2006dip}.
In a $q \bar{q} + p$ scattering, $r$ is the size of the $q \bar{q}$ dipole and
corresponds to the spatial scale of the probe.

The positions of nucleons inside a nucleus are sampled according to
the Wood-Saxon distribution and
$\textbf{x}_{T,i}$ is the position of the $i$-th nucleon
in the transverse plane.
Once the color charge distribution is determined,
we solve the classical Yang-Mills equation to obtain the gluon field in
each nucleus
\begin{equation}
	A_{(1,2)}^{i} = - \frac{i}{g} U_{(1,2)}
		\partial_i U_{(1,2)}^{\dagger} (\textbf{x}_T)
\end{equation}
where the Wilson line $U$ is
\begin{equation}
	U_{(1,2)} (\textbf{x}_T) = \mathcal{P} \exp{ \left[
		-ig \int dx^{(+,-)}
		\frac{\rho_{(1,2)} (\textbf{x}_T, x^{(+,-)}) }{\nabla_T^2 - m^2}
		\right] } \, .
\label{eq:U12}
\end{equation}
The subscript $(1)$ and $(2)$ indicates projectile and target quantities,
respectively. 
Since the fluctuation scale of $\rho_{(1,2)}$ is $\sim Q_s$,
so is the fluctuation scale in $A^i_{(1,2)}$.

The gluon field right after the collision ($\tau \to 0^+ $) is
given by~\cite{Krasnitz:1999wc,Krasnitz:2000gz,Krasnitz:2002mn}
\begin{eqnarray}
	A^i ( \tau \to 0^+ ) & = & A_{(1)}^i + A_{(2)}^i \\
	A^{\eta} ( \tau \to 0^+ ) & = & \frac{ig}{2} [ A_{(1)}^i , A_{(2)}^i ] \, .
\end{eqnarray}
For $\tau > 0$,
we evolve the gluon field according to the Yang-Mills equation
\begin{equation}
	\partial_{\mu} F^{\mu\nu} - ig [ A_{\mu}, F^{\mu\nu} ]
		= 0
\end{equation}
where the field strength tensor is given as usual by
\begin{equation}
	F^{\mu\nu} = \partial^{\mu} A^{\nu} - \partial^{\nu} A^{\mu}
		- ig [ A^{\mu}, A^{\nu} ]\;.
\end{equation}
After evolving the gluon field up to $\tau_0 = 0.4 \,\textrm{fm}$,
the energy-momentum tensor is formed out of the field strength tensor
\begin{equation}
	T^{\mu\nu} = - 2 \,\textrm{Tr} \,
		\left( F^{\mu}_{\phantom{\mu}\alpha} F^{\nu\alpha} \right)
		+ \frac{1}{2} g^{\mu\nu}
		\textrm{Tr} \, \left(F^{\alpha\beta} F_{\alpha\beta} \right)
\end{equation}
where the trace is over color in the fundamental representation.
The time-like eigenvalue of $T^{\mu}_{\ \nu} = T^{\mu\lambda}g_{\lambda\nu}$
provides the local energy density and the flow velocity
\begin{equation}
	T^{\mu}_{\phantom{\mu}\nu} u^{\nu} = \epsilon \, u^{\mu}
\end{equation}
of IP-Glasma at $\tau_0$. 

The normalization of the energy-momentum tensor in IP-Glasma
is not fully constrained, owing to freedom in the choice of $\alpha_s$,
as noted in Ref.~\cite{Schenke:2012fw}.
This normalization can be fixed
by comparing the results of the hydrodynamical simulation
with charged hadron multiplicity measurements.
This is the procedure adopted in this work.
Effectively, this translates into a normalization of the energy density
$\epsilon$ of IP-Glasma.
The initial flow $u^{\mu}$ is unaffected by this normalization of $T^{\mu\nu}$.
Note that the shear stress tensor of IP-Glasma is not currently used
to initialize the hydrodynamic simulation,
where $\pi^{\mu\nu}$ is initialized to zero.

\subsection{Second-order relativistic hydrodynamics
with shear and bulk viscosity}

The main hydrodynamic equations are the conservation laws
of net-charge, energy, and momentum.
Since we only aim to describe the matter produced in the mid-rapidity region
at high collision energies,
the net-charge can be approximated to be zero
and we are only required to solve the continuity equation for $T^{\mu \nu }$, 
\[
\partial _{\mu }T^{\mu \nu }=0.
\]

In a viscous fluid, the energy-momentum tensor $T^{\mu \nu }$ is decomposed
in terms of the velocity field as 

\begin{equation}
	T^{\mu\nu} = \epsilon \, u^{\mu} u^{\nu}
		- \left( P + \Pi \right) \Delta^{\mu\nu}
		+ \pi^{\mu\nu}
\end{equation}
where $\Delta ^{\mu \nu }\equiv g^{\mu \nu }-u^{\mu }u^{\nu }$,
$P$ is the thermodynamic pressure,
$\Pi $ is the bulk viscous pressure,
and $\pi ^{\mu \nu }$ is the shear stress tensor.
The relation between $\epsilon $ and $P$ is given
by an equation of state, $P(\epsilon )$.
In this work we use the equation of state constructed
from a hadronic resonance gas and lattice calculation \cite{Huovinen:2009yb}.

\begin{figure}[h]
\begin{center}
\includegraphics[width=0.45\textwidth]{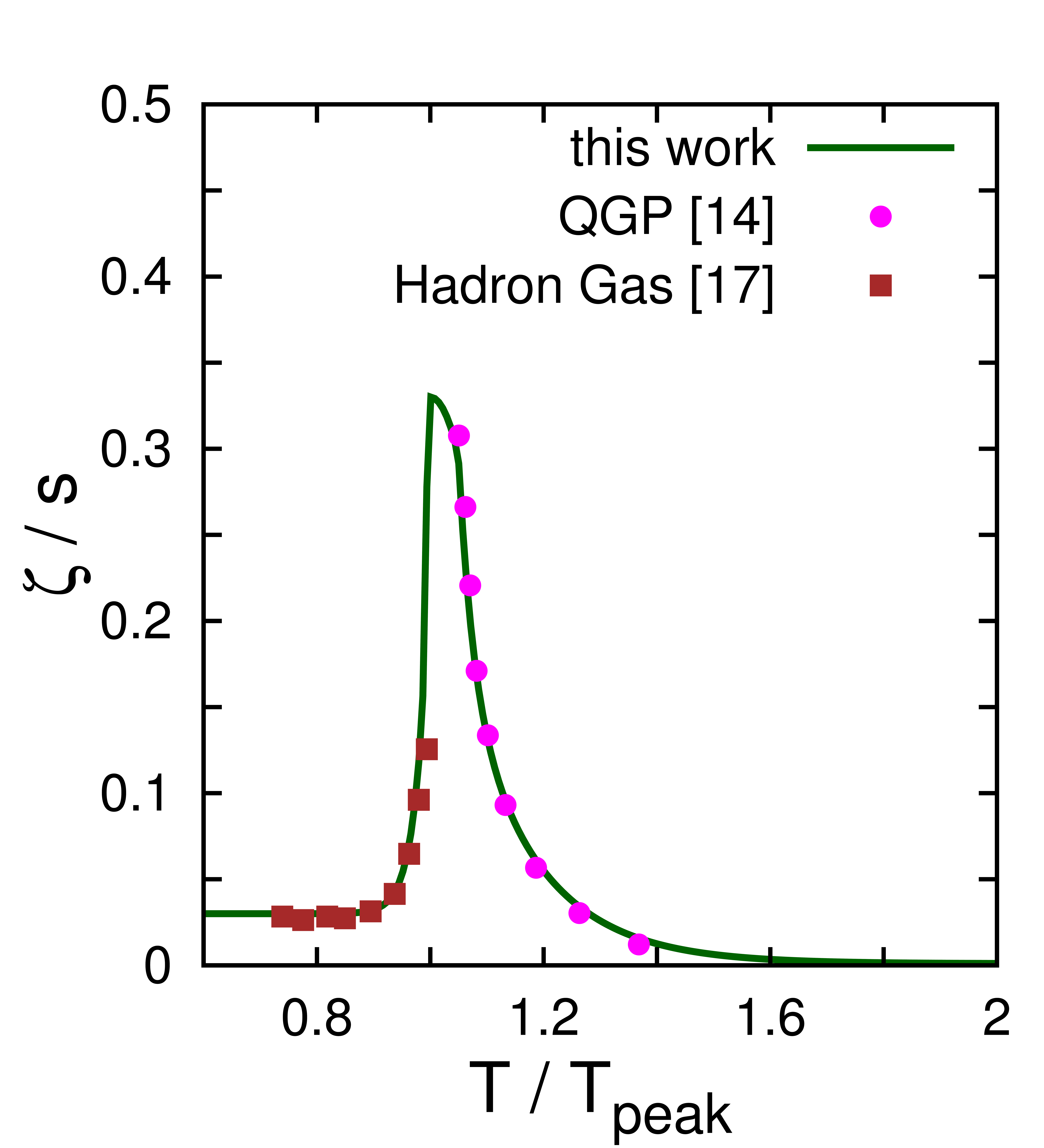}
\end{center}
\caption{
The temperature dependence of the bulk viscosity over entropy density ratio
used in this study.
The QGP side of the $\zeta/s$ is taken from Ref.~\cite{Karsch:2007jc}
and the hadronic side is taken from Ref.\cite{NoronhaHostler:2008ju}.
}
\label{fig:zeta_over_s}
\end{figure}

The time evolution of the bulk and shear viscous corrections,
driven by the expansion rate $\theta = \nabla_{\mu} u^{\mu}$
and the shear tensor $\sigma^{\mu\nu}
= \frac{1}{2}\left[ \nabla^{\mu} u^{\nu} + \nabla^{\nu} u^{\mu}
- \frac{2}{3} \Delta^{\mu\nu} (\nabla_{\alpha} u^{\alpha}) \right]$,
in which $\nabla_{\mu} = ( g_{\mu\nu} - u_{\mu} u_{\nu} ) \partial^{\nu}$,
are given by the equations
\begin{eqnarray}
	\tau_{\Pi} \dot{\Pi} + \Pi
	& = & - \zeta\, \theta
	- \delta_{\Pi \Pi} \Pi \, \theta
	+ \lambda_{\Pi \pi} \pi^{\mu\nu} \sigma_{\mu\nu} \\
	\tau_{\pi} \dot{\pi}^{\langle \mu\nu \rangle}
	+ \pi^{\mu\nu}
	& = & 2\eta\,\sigma^{\mu\nu}
	- \delta_{\pi \pi} \pi^{\mu\nu} \theta
	+ \varphi_7 \pi_{\alpha}^{\langle \mu} \pi^{\nu\rangle \alpha} \nonumber \\
	& & - \tau_{\pi \pi} \pi_{\alpha}^{\langle \mu} \sigma^{\nu\rangle \alpha}
	+ \lambda_{\pi \Pi} \Pi\, \sigma^{\mu\nu}
\end{eqnarray}
which follows from 
the 14-moment approximation of the Boltzmann equation~
\cite{Denicol:2012cn,Denicol:2014vaa}.
The first-order transport coefficients $\eta$ and $\zeta$ are
the shear and bulk viscosities, respectively.
The shear and bulk relaxation time, $\tau_{\pi}$ and $\tau_{\Pi}$ are
found to be
\begin{eqnarray}
	\tau_{\pi} & = & \frac{5 \, \eta}{\epsilon + P } \, , \\
	\tau_{\Pi} & = & \frac{\zeta}{15 \,
		\left( \frac{1}{3} - c_s^2 \right)^2 ( \epsilon + P ) } 
\end{eqnarray}
where $c_s$ is the speed of sound.
The second-order transport coefficients are
related to the relaxation time through the relations
\begin{eqnarray}
	\frac{ \delta_{\pi \pi} }{ \tau_{\pi} } & = & \frac{4}{3} \\
	\frac{ \tau_{\pi \pi} }{ \tau_{\pi} } & = & \frac{10}{7} \\
	\frac{ \lambda_{\pi \Pi} }{ \tau_{\pi} } & = & \frac{6}{5} \\
	\frac{ \delta_{\Pi \Pi} }{ \tau_{\Pi} } & = & 1 - c_s^2 \\
	\frac{ \lambda_{\Pi \pi} }{ \tau_{\Pi} } & = &
		\frac{8}{5} \left( \frac{1}{3} - c_s^2 \right) \\
	\varphi_7 & = & \frac{18}{35} \frac{1}{ \epsilon + P }\;.
\end{eqnarray}
These values were first obtained in \cite{Denicol:2014vaa}.
Once we have $\eta$ and $\zeta$ as functions of temperature,
it is possible to find the temperature dependence of the relaxation times
and the remaining second-order transport coefficients.
The shear viscosity over the entropy density ratio $\eta/s$ is set
to be constant in this work.
Data-driven determination of the temperature-dependent $\eta/s$
via Bayesian analysis is performed in \cite{Bernhard:2016tnd}.
The temperature dependence of the 
bulk viscosity to entropy density ratio
$\zeta/s$ is fixed as shown in Fig.~\ref{fig:zeta_over_s}
and the temperature where bulk viscosity peaks is set
to be $T_{\scriptsize \textrm{peak}} = 180\,\textrm{MeV}$
based on the transition temperature in the equation of state.

\subsection{Transition from hydrodynamics to transport theory}

During the hydrodynamic evolution,
the system becomes gradually more dilute and,
at some point, a hydrodynamic description will break down.
Nevertheless the system is still interacting and
the subsequent dynamics must be described in another framework,
transport theory for example.
Using transport generally means that fluid elements must be converted
into hadronic degrees of freedom,
which will then be described using a hadronic kinetic theory simulation.
In principle, this matching between degrees of freedom must be performed
in a space-time region in which both hydrodynamics and kinetic theory
are within their domain of applicability. 

One way to transition from hydrodynamics to transport theory is
to do so when the expansion rate of the fluid is
moderately (but not significantly) smaller
than the mean-free path of the hadrons composing the fluid.
In this way, the system is interacting enough for hydrodynamics
to apply and dilute enough for the Boltzmann equation to be applicable.
In practice, implementing such procedure can be rather complicated,
since it requires extensive knowledge about the interactions among the hadrons.
In this work, we adopt the common simplification of
approximating the switching hypersurface
as a constant temperature hypersurface,
with the switching temperature $T_{\rm sw}$
becoming one of the free parameters of our model.
As we shall discuss in the following sections,
this parameter will be determined by optimizing
the fit of identified charged hadron multiplicity, especially protons. 

It is important to emphasize
that $T_{\rm sw}$ is an effective parameter of the model,
that is supposed to describe the more complicated physics of the transition
from the description of dense to dilute systems.
In this sense, there is no reason to expect such temperature
to remain the same as one changes the collision energy (from RHIC to LHC)
or even centrality class.
One of our conclusions will be
that the switching temperature at RHIC (=165 MeV) is larger
than the switching temperature at the LHC (=145 MeV).
This difference may be due to the fact that systems produced at LHC energies
have more entropy and, consequently, are more long-lived
than the ones produced at RHIC.
Naturally, a more precise explanation for the switching parameters
we extract can only be obtained by improving the model,
taking into account a more realistic transitioning to the transport phase
with improved viscous correction to the distribution function.

We now present the details of how we switch
from hydrodynamics to hadronic transport.
On isothermal hypersurfaces with constant switching temperature 
$T_{\rm sw}$,
we sample particles with degeneracy $d$ and mass $m$
according to the Cooper-Frye formula \cite{Cooper:1974mv}
\begin{eqnarray}
	\frac{dN}{d^3 \textbf{p}} & = & \frac{d}{(2\pi)^3} \int_{\Sigma}
		\frac{p^{\mu} d^3 \Sigma_{\mu}}{E_{\scriptsize \textbf{p}}} \nonumber\\
	& & \times \left[ f_0 (x,\textbf{p})
		+ \delta f_{\scriptsize \textrm{shear}} (x,\textbf{p})
		+ \delta f_{\scriptsize \textrm{bulk}} (x,\textbf{p})
		\right] \hspace{10pt}
	\label{eq:CooperFrye}
\end{eqnarray}
where $E_{\scriptsize \textbf{p}}$ satisfies
$E_{\scriptsize \textbf{p}}^2 = \textbf{p}^2 + m^2$.
The normal vector $d^3 \Sigma_{\mu}$
is an exterior product of three displacement vectors
tangential to the hypersurface.
In our simulation, we construct the hypersurface from tetrahedra \cite{MUSIC}
and sample hadrons at each of the grid locations $x$.
We approximate the probability distribution of the number of particles to be
a Poisson distribution whose average value is given by
\begin{equation} \begin{array}{l}
	\displaystyle \bar{N} |_{\scriptsize \textrm{1-cell}} \vspace{2pt}\\
	\displaystyle = \left\{ \begin{array}{ll}
		\displaystyle \left[ n_0 (x)
		+ \delta n_{\scriptsize \textrm{bulk}} (x) \right]
		u^{\mu} \Delta \Sigma_{\mu}
		& \quad \textrm{if} \quad u^{\mu} \Delta \Sigma_{\mu}
		\ge 0 \vspace{2pt}\\
		0 & \quad \textrm{otherwise}
		\end{array}
		\right.
\end{array} \end{equation}
where the number density at thermal equilibrium $n_0$
and the bulk viscous correction $\delta n_{\scriptsize \textrm{bulk}}$
are given by
\begin{eqnarray}
	n_0 (x) & = & d \int \frac{d^3 \textbf{k}}{(2\pi)^3} \,
		f_0 ( x, \textbf{k} ) \\
	\delta n_{\scriptsize \textrm{bulk}} (x)
		& = & d \int \frac{d^3 \textbf{k}}{(2\pi)^3} \,
		\delta f_{\scriptsize \textrm{bulk}} ( x, \textbf{k} ) \;.
\end{eqnarray}
The shear tensor correction does not induce a change in the number density
because of its spin-2 structure (see Eq.(\ref{eq:dfshear}) below),
which is orthogonal to any scalar.

It is understood that quantum thermal distributions are not Poissonian, since
\begin{eqnarray}
	\langle N^2 \rangle - \langle N \rangle^2
		& = & d \, V \int \frac{d^3 \textbf{k}}{(2\pi)^3} \,
		f_0 (\textbf{k}) \, \left( 1 \pm f_0 (\textbf{k}) \right)
		\hspace{10pt}\\
	& \ne & \langle N \rangle\;.
\end{eqnarray}
Nevertheless, 
within the range of switching temperature
considered in this work ($135\,\hbox{MeV} \le T_{\rm sw} \le 165\,\hbox{MeV}$),
we verified that these quantum effects are less than $10\,\%$ for pions and
less than $1\,\%$ for heavier hadrons.
Using the Poisson distribution is therefore a reasonable approximation.

After we determine the number of particles in each cell,
we sample the momentum of each particle
according to the following prescription~\cite{Huovinen:2012is}
\begin{equation}
	\left. \frac{dN}{d^3 \textbf{p}} \right|_{\scriptsize \textrm{1-cell}}
		= \frac{d}{(2\pi)^3} \left[ f_0+ \delta f_{\scriptsize \textrm{shear}} 
		+ \delta f_{\scriptsize \textrm{bulk}}
		\right] \frac{p^{\mu} \Delta \Sigma_{\mu}}{E_{\scriptsize \textbf{p}}}
	\label{eq:CooperFrye_1Cell}
\end{equation}
if $(f_0 + \delta f_{\scriptsize \textrm{shear}}
+ \delta f_{\scriptsize \textrm{bulk}}) > 0$
and $p^{\mu} \Delta \Sigma_{\mu} > 0$.
Otherwise, $	\left. dN/d^3 \textbf{p} \right|_{\scriptsize \textrm{1-cell}}=0$.
In our simulation, the deviation from particle spectra
given by Riemann integration of Eq.(\ref{eq:CooperFrye})
is less than 5\% for $p_T < 2\,\textrm{GeV}$.
Therefore, as long as the soft physics is concerned,
Eq.(\ref{eq:CooperFrye_1Cell}) is an adequate implementation
of the Cooper-Frye transition.

The explicit expressions for the equilibrium distribution functions
and the shear \cite{Dusling:2010rel}
and bulk \cite{Bozek:2009dw,Paquet:2015lta} viscous corrections are
\begin{eqnarray}
	f_0 & = & \frac{1}{\exp{(p\cdot u/T)} \mp 1} \\
	\delta f_{\scriptsize \textrm{shear}}
		& = & f_0 (1 \pm f_0)
		\frac{\pi_{\mu\nu} p^{\mu} p^{\nu}}{2\, (\epsilon_0 + P_0) T^2}
		\label{eq:dfshear}
		\\
	\delta f_{\scriptsize \textrm{bulk}}
		& = & - f_0 (1 \pm f_0) \frac{ C_{\scriptsize \textrm{bulk}} }{T}
		\nonumber\\
		& & \times \left[
		\frac{m^2}{3 \, (p \cdot u)}
		- \left(\frac{1}{3} - c_s^2 \right) (p \cdot u)
		\right] \Pi
\label{eq:dfbulk}
\end{eqnarray}
where
\begin{eqnarray}
	\frac{1}{ C_{\scriptsize \textrm{bulk}} }
		& = & \frac{1}{3T} \sum_n d_n m_n^2
		\int \frac{d^3 \textbf{k}}{(2\pi)^3 E_{\scriptsize \textbf{k}}}
		\nonumber\\
		& & \times f_{n,0} \, (1 \pm f_{n,0} ) \left[
		\frac{m_n^2}{3 E_{\scriptsize \textbf{k}} }
		- \left(\frac{1}{3} - c_s^2 \right) E_{\scriptsize \textbf{k}}
		\right] \hspace{10pt}
\end{eqnarray}
and the flow velocity $u^{\mu}$ and temperature $T$ on the hypersurface
are determined from the hydrodynamic evolution.
The summation is over hadronic species.

\subsection{Microscopic transport \texttt{UrQMD} as afterburner}

The sampled particles
are propagated in \texttt{UrQMD} (version 3.4) \cite{UrQMD1,Bleicher:1999xi},
which simulates interactions of hadrons and resonances
with masses up to $2.25\,\textrm{GeV}$.
These interactions include inelastic processes through resonance scattering,
$B\bar{B}$ annihilation and string excitation, as well as elastic scatterings.
Whenever experimental data are available,
the hadronic cross sections in \texttt{UrQMD} are based on the data.
When measurements are not available,
cross sections are extrapolated from other processes
based on detailed balance and the additive quark model.
Using \texttt{UrQMD} as afterburner allows for a more realistic description
of the late stage of the collision, 
where the mean free path is not short compared to the macroscopic scale
given by system size or expansion rate.

Note that while baryon-antibaryon annihilation is included in UrQMD,
pair creation is not. 
This is because $B\bar{B}$ predominantly annihilate into multiple pions,
but the opposite channel,
which would involve the simultaneous interaction of multiple hadrons,
is not currently supported by UrQMD.
Consequently, all $B\bar{B}$ pairs in the
system originate from the Cooper-Frye procedure and,
strictly speaking, detailed balance is not obeyed.
This violation of detailed balance is not expected to be a major issue:
previous works such as Ref.~\cite{Pan:2014caa} have shown
that the contribution of baryon-antibaryon creation
is considerably smaller than that of $B\bar{B}$ annihilation.
It is also possible to make the argument
that since the system is expanding and
the mean free path is comparable to the macroscopic scale,
there are more baryons and anti-baryons
than there would be in local thermal equilibrium.
This is because the system does not have enough interactions
to reach equilibrium.
The excess of mesons over thermal equilibrium is less significant
owing to the lower masses.
Therefore, one can expect
that $B\overline{B}$ annihilation will be more frequent
than the inverse process in the evolution toward equilibrium.
In this sense,
switching from hydrodynamics to transport coincides with
the point where the $B\bar{B}$ annihilation becomes dominant.

\section{Results and discussion}
\label{sec:results}

In this section, we discuss the results of our simulations
for Au-Au collisions at RHIC ($\sqrt{s_{NN}}=200$~GeV) and
Pb-Pb collisions at the LHC ($\sqrt{s_{NN}}=2.76$~TeV)
where a wide set of measurements are available.
The centrality classes
$0-5\%$, $10-20\%$, $20-30\%$ and $30-40\%$ are considered.
We highlight the effect of bulk viscosity 
and the importance of the hadronic rescattering stage.
We show that our approach
is capable of describing a large number of hadronic observables consistently
with a fixed set of parameters.
The main parameters in this work are the switching temperature between
the hydrodynamic expansion and the afterburner \texttt{UrQMD}, $T_{\rm sw}$,
and the value of the effective shear viscosity over the entropy density ratio
$\eta/s$.

\subsection{Integrated observables}
\label{sec:integrated}

\begin{figure}[h]
\begin{center}
\includegraphics[width=0.49\textwidth]{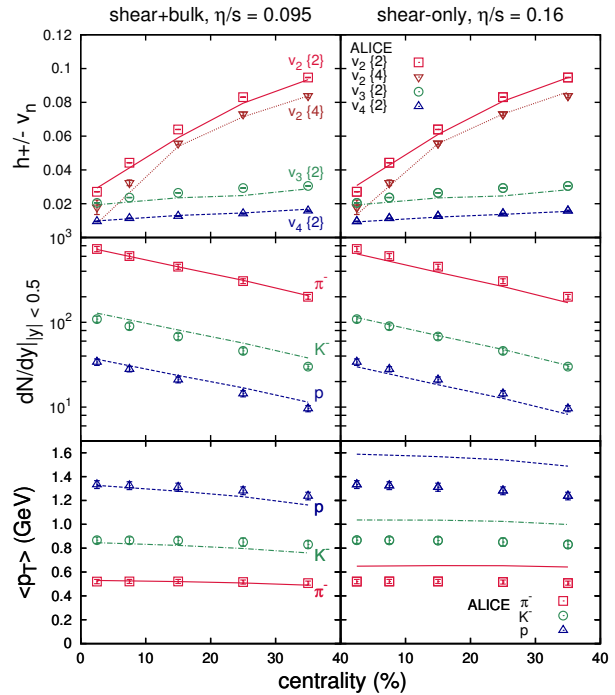}
\caption{Integrated $v_n$ (upper),
mid-rapidity multiplicity $dN/dy|_{y=0}$ (middle)
and mean-$p_T$ (lower) as functions of centrality.
The ratio of the shear viscosity to entropy density $\eta /s$ is determined
to fit the ALICE data on $v_n$ \cite{ALICE:2011ab}.
The left-hand panels include bulk viscosity
as shown in Fig.~\ref{fig:zeta_over_s},
while the right-hand panels were computed with $\zeta/s=0$.
The non-zero bulk viscosity alters the favored value of $\eta /s$.
The ALICE data \cite{Abelev:2013vea}
for $dN/dy|_{|y|<0.5}$ and $\langle p_T \rangle$ are also shown.
}
\label{dNdy_pTave_vnint_Ttr145}
\end{center}
\end{figure}

\begin{figure}[h]
\begin{center}
\includegraphics[width=0.49\textwidth]{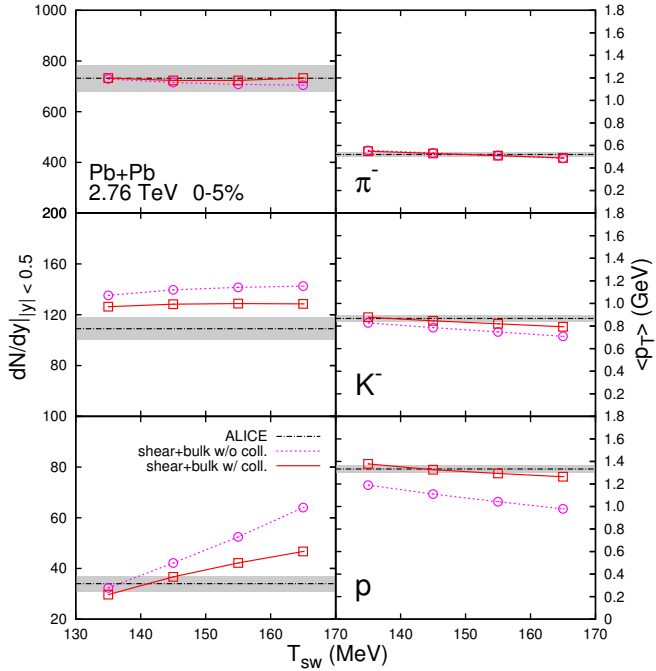}
\caption{Mid-rapidity multiplicity $\left. dN/dy \right|_{ |y| < 0.5 }$ (left)
and mean-$p_T$ (right) of pions, kaons and protons
as functions of the switching temperature $T_{\scriptsize \textrm{sw}}$.
The most central Pb-Pb collisions with $\sqrt{s_{NN}} = 2.76\,\textrm{TeV}$
are considered.
The ALICE data \cite{Abelev:2013vea} are shown as the bands.}
\label{dNdy_pTave_vsTtr_0005}
\end{center}
\end{figure}

\begin{figure}[h]
\begin{center}
\includegraphics[width=0.49\textwidth]{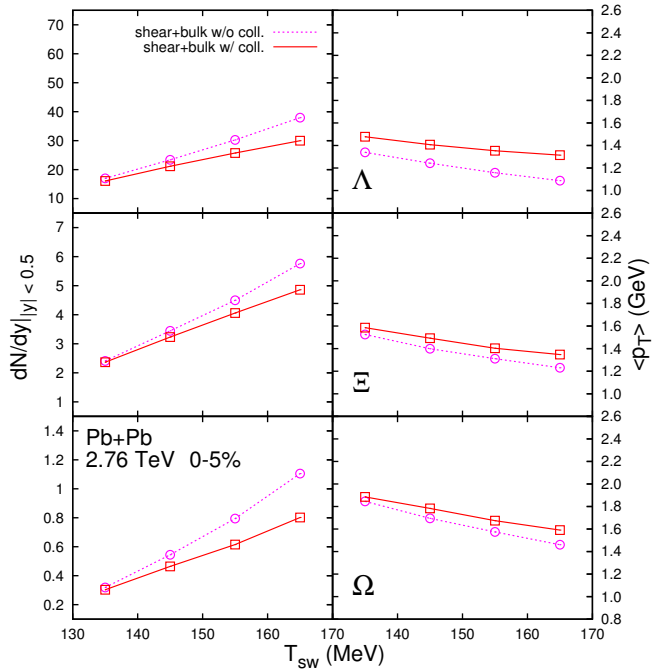}
\caption{Mid-rapidity multiplicity $\left. dN/dy \right|_{ |y| < 0.5 }$ (left)
and mean-$p_T$ (right) of $\Lambda$, $\Xi$ and $\Omega$ baryons
as functions of the switching temperature $T_{\scriptsize \textrm{sw}}$.
The most central Pb-Pb collisions with $\sqrt{s_{NN}} = 2.76\,\textrm{TeV}$
are considered.
}
\label{dNdy_pTave_str_vsTtr_0005}
\end{center}
\end{figure}

\begin{figure}[h!]
\begin{center}
\includegraphics[width=0.49\textwidth]{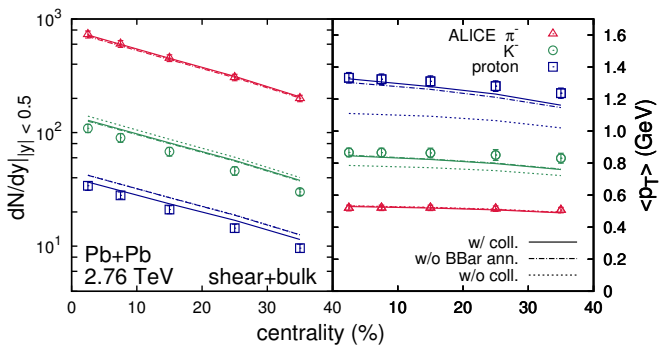}
\end{center}
\caption{Mid-rapidity multiplicity (left panel) and mean $p_T$ (right panel)
of identified particles as functions of centrality.
}
\label{dndy_ave_pt_with_bulk_cent}
\end{figure}

Observables integrated over the transverse momentum $p_T$
generally have a reduced sensitivity
to out-of-equilibrium corrections of the hadronic momentum distribution
($\delta f_{\scriptsize \textrm{shear}}$) because of shear viscosity,
compared to $p_T$-differential observables.
Therefore, the multiplicity $\left. dN/dy \right|_{ |y| < 0.5 }$,
the mean transverse momentum $\langle p_T \rangle$,
and the $p_T$-integrated anisotropic flow coefficients $v_n$
are investigated first.
The anisotropic flow $v_n$ is computed using the multi-particle cumulant method
based on the flow correlations among particles
as in Ref.~\cite{Bilandzic:2010jr}.

Figure~\ref{dNdy_pTave_vnint_Ttr145} shows
the multiplicity and average $p_T$ for pions, kaons and protons,
as well as the charged hadron anisotropic flow coefficients $v_{2,3,4}$,
from central to semi-peripheral centrality bins.
The charged hadron $v_2$ is shown for the two- and four- particle cumulants.
Calculations that include both shear and bulk viscosities
are in the left panels,
while the calculations with only shear viscosity are presented
in the right panels.
In both cases, the value of $\eta/s$ was adjusted
such that the measured charged hadron $v_n$ is reproduced.
A value of $\eta/s=0.095$ is used
when both bulk and shear viscosity are present,
while a larger value of $\eta/s=0.16$ is necessary
in absence of bulk viscosity.
This important effect of bulk viscosity
on phenomenological extractions of $\eta/s$ had been quantified previously
in Ref.~\cite{Ryu:2015vwa}. 

Another significant effect of bulk viscosity is a considerable suppression
of the average transverse momentum of hadrons,
which can be seen by comparing the left and right hand sides
of Fig.~\ref{dNdy_pTave_vnint_Ttr145}.
The change in average $p_T$ is significant for all hadron species
(pions, kaons, protons).
Bulk viscosity is essential for a simultaneous description
of the multiplicity and $\langle p_T \rangle$ of hadrons
when IP-Glasma initial conditions are used:
without bulk viscosity, the system expands too rapidly,
leading to a larger hydrodynamic transverse flow
than suggested by average $p_T$ measurements.
Bulk viscosity improves the agreement with data
by acting as a resistance to expansion,
reducing the transverse flow of the system.
Besides this change in the plasma expansion,
part of the modification of $\langle p_T \rangle$ is
from the effect of bulk viscosity on the hadronic momentum distribution ---
$\delta f_{\scriptsize \textrm{bulk}}$ given by Eq.(\ref{eq:dfbulk}).
The average $p_T$ is actually
decreased by $\delta f_{\scriptsize \textrm{bulk}}$.
If $\delta f_{\scriptsize \textrm{bulk}}$ were smaller,
a similar suppression in $\langle p_T \rangle$ could be achieved
with a larger bulk viscosity.
More details about the effect of $\delta f_{\scriptsize \textrm{bulk}}$
on integrated hadronic observables are presented
in Appendix~\ref{sec:appendixdf}. 
We highlight here
that $\delta f_{\scriptsize \textrm{bulk}}$ has a small effect
on the $v_n$ of charged hadrons and the multiplicity of pions and kaons.

The switching temperature $T_{\rm sw}$
between the hydrodynamic simulation and the hadronic afterburner is
$145$~MeV for the calculations presented in Fig.~\ref{dNdy_pTave_vnint_Ttr145}.
This choice can be understood
from Fig.~\ref{dNdy_pTave_vsTtr_0005} (solid lines),
which shows the dependence on $T_{\rm sw}$ of the multiplicity
and average transverse momentum of identified hadrons
(the charged hadron momentum anisotropies have a small dependence
on $T_{\rm sw}$ --- this is discussed in more details below).
The multiplicities of pions and kaons are shown
to have a weak dependence on $T_{\rm sw}$,
while protons are much more sensitive to this parameter.
The value of $T_{\rm sw}$ around $145$~MeV leads to
the best agreement with ALICE measurements for the proton multiplicity. 

Figure~\ref{dNdy_pTave_vsTtr_0005} (solid lines) also shows
that dependence of the average transverse momentum
on $T_{\rm sw}$ is mild for all three hadrons species.
We verified that a similar $T_{\scriptsize \textrm{sw}}$ dependence
was found for more peripheral collisions, up to the $30-40\%$ centrality class,
for both the multiplicity and $\langle p_T \rangle$.
We further verified that the dependence on the switching temperature
$T_{\scriptsize \textrm{sw}}$ of these same observables is very similar
with and without bulk viscosity.
The effect of $T_{\scriptsize \textrm{sw}}$
on the identified hadron $\langle p_T \rangle$ is thus small
compared to the effect of bulk viscosity.
In consequence, we emphasize that in absence of bulk viscosity,
it would not be possible to obtain a good agreement
with identified hadron $\langle p_T \rangle$
by changing the value of $T_{\scriptsize \textrm{sw}}$.

Also shown in Fig.~\ref{dNdy_pTave_vsTtr_0005} is
how hadronic rescattering affects the $T_{\scriptsize \textrm{sw}}$ dependence
of the multiplicity and average transverse momentum.
It is found that calculations that include hadronic decays
but not hadronic interactions (dashed lines) have a larger slope
in $T_{\scriptsize \textrm{sw}}$
than those that include both hadronic decays and rescattering (solid lines).
This means that hadronic rescattering reduces the dependence
on $T_{\scriptsize \textrm{sw}}$.
The effect is fairly small for pions and kaons, but significant for protons.
Heavier baryons, shown in Fig.~\ref{dNdy_pTave_str_vsTtr_0005},
have a similar dependence on $T_{\scriptsize \textrm{sw}}$ as protons.
Since hydrodynamics describes an interacting medium,
it is indeed expected that the transition between hydrodynamics and UrQMD
will be smoother --- if not necessarily smooth ---
when hadronic rescattering is included.
The larger dependence of protons and heavier hadrons
on $T_{\scriptsize \textrm{sw}}$ can be seen as a systematic uncertainty
of our model for observables involving these hadrons.

The effect of hadronic rescattering
on our multiplicity and mean $p_T$ calculations is shown again
in Fig.~\ref{dndy_ave_pt_with_bulk_cent},
this time as a function of the centrality class,
and compared against ALICE data.
The switching temperature is fixed $T_{\scriptsize \textrm{sw}}=145$~MeV,
which as explained above provides a good description
of the proton multiplicity.
It can be seen in Fig.~\ref{dndy_ave_pt_with_bulk_cent}
that the effect of hadronic rescattering is very similar across centralities.
We also show in Fig.~\ref{dndy_ave_pt_with_bulk_cent}
the explicit effect of a subset of hadronic rescatterings,
namely baryon-antibaryon annihilation.
What is interesting about $B\bar{B}$ annihilations is
that they represent most of the change in the multiplicity of protons
due to hadronic rescattering,
although they have a very small effect
on the average transverse momentum of protons.
This result is in general agreement with the observations
made in \cite{Song:2013qma}.

\begin{figure}[tbp]
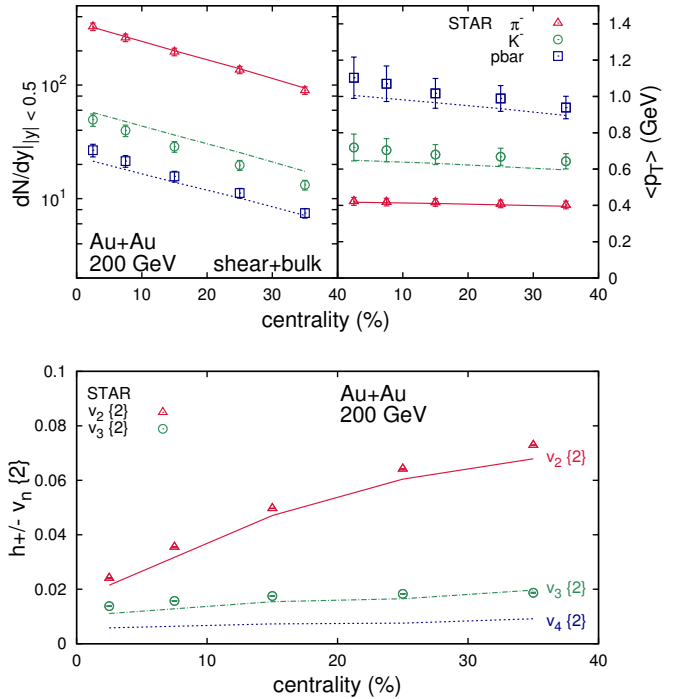

\begin{center}
\includegraphics[width=0.49\textwidth]{dNdy_pTave_WBulk_tune3_Ttr165.pdf}
\hspace{0.5cm}
\includegraphics[width=0.49\textwidth]{vnint_WBulk_tune3_Ttr165.pdf}
\caption{(a) Mid-rapidity multiplicity $\left. dN/dy \right|_{ |y| < 0.5 }$ 
and mean-$p_T$ of pions, kaons and protons and
(b) charged hadron momentum anisotropy, as functions of centrality,
for Au-Au collisions with $\sqrt{s_{NN}} = 200\,\textrm{GeV}$.
The value of $T_{\scriptsize \textrm{sw}}$ is set to $165$~MeV
so as to provide the a good description of the proton multiplicity,
while $\eta /s = 0.06$ was adjusted to describe the momentum anisotropies.
Measurements from STAR \cite{
Adams:2004bi,Abelev:2008ab,Adamczyk:2013waa} are also shown.
}
\label{fig:RHICintegrated}
\end{center}
\end{figure}

\begin{figure}[tbp]
\begin{center}
\includegraphics[width=0.49\textwidth]{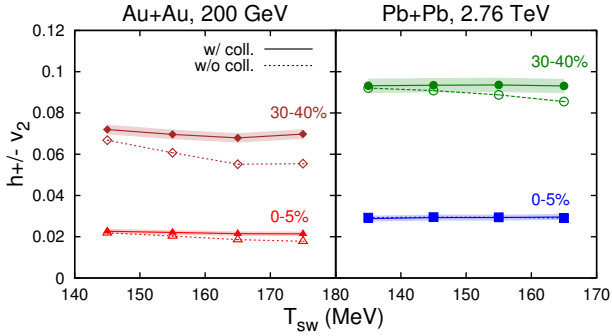} 
\caption{(a) Mid-rapidity charged hadron $v_2\{2\}$ with
and without hadronic rescattering,
as a function of $T_{\scriptsize \textrm{sw}}$ for Au-Au collisions
with $\sqrt{s_{NN}} = 200\,\textrm{GeV}$ (left) and
for Pb-Pb collisions with $\sqrt{s_{NN}} = 2.76\,\textrm{TeV}$ (right).
}
\label{fig:RHIC_LHC_vn_Tsw}
\end{center}
\end{figure}

In Fig.~\ref{fig:RHICintegrated},
we show the results of our calculations at RHIC by comparing
with STAR measurements~\cite{Adams:2004bi,Abelev:2008ab,Adamczyk:2013waa}.
Like it was done with LHC calculations,
the value of $T_{\rm sw}$ was adjusted
so as to provide a good description of the proton multiplicity,
while the value of $\eta/s$ was fixed using the charged hadron $v_2$.
The values of $T_{\rm sw}=165$~MeV and $\eta/s=0.06$
were found to provide good agreement with the respective measurements.
The quality of agreement with measurements can be seen to be similar to
that found at the LHC (c.f. Fig.~\ref{dNdy_pTave_vnint_Ttr145}).
We verified that the dependence on $T_{\scriptsize \textrm{sw}}$
of the multiplicity and average transverse momentum of identified hadrons is
very similar at RHIC as at the LHC.

\begin{figure}[b]
\begin{center}
\includegraphics[width=0.49\textwidth]{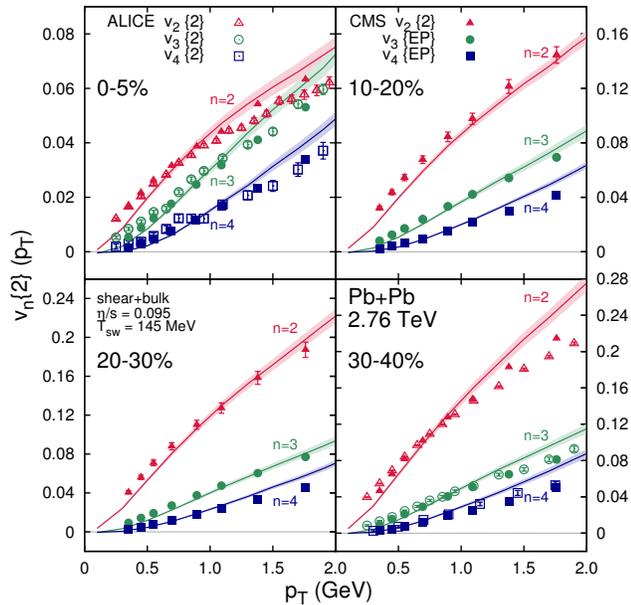}
\caption{
$p_T$ differential $v_n \{ 2 \}$ ($n = 2,\, 3 \,\,\textrm{and} \,\, 4$)
of charged hadrons
for centrality classes $0-5\%$, $10-20\%$, $20-30\%$ and $30-40\%$
of Pb-Pb collisions
with $\sqrt{s_{\scriptsize \textrm{NN} } } = 2.76\,\textrm{TeV}$.
The statistical errors in the calculation are shown
as the bands around the curves.
The ALICE data \cite{ALICE:2011ab} and CMS \cite{CMS:2013v2,CMS:2014vn} data
are also shown for comparison.
}
\label{vnpT_hch_LHC}
\end{center}
\end{figure}

To conclude this section on integrated observables,
in Fig.~\ref{fig:RHIC_LHC_vn_Tsw} we investigate
the effect of the switching temperature $T_{\scriptsize \textrm{sw}}$
between hydrodynamics and UrQMD
on the momentum anisotropy $v_2\{2\}$ of integrated charged hadrons,
at RHIC as well as at the LHC.
The upper curves correspond to peripheral 30-40\% collisions,
and the lower curves to central 0-5\% collisions.
The solid line corresponding to the calculation
with hadronic decays and rescattering,
and the dashed line including only hadronic decays but not rescattering.
As observed previously for the multiplicity and average transverse momentum,
the inclusion of hadronic rescattering reduces significantly
the observable's dependence on $T_{\scriptsize \textrm{sw}}$
(i.e. solid lines flatter than dashed ones).
At both RHIC and LHC, hadronic rescattering increases $v_n$,
which is consistent with the effect of rescattering observed on pions
in Ref.~\cite{Hirano:2007ei}.
This increase is larger at RHIC than at the LHC,
and is also larger for peripheral events than for central ones.
Our understanding is that this is a consequence of the different lifetime
of the hadronic transport phase compared to the hydrodynamic expansion
for the different centralities and collision energies,
as well as a consequence of how isotropic each system is
at the transition between hydrodynamics and transport.

\subsection{Differential observables}

In this section, we examine $p_T$ differential observables.
At this point, all model parameters have already been fixed
with integrated observables both at LHC and RHIC energies.

\begin{figure}[tbp]
\begin{center}
\includegraphics[width=0.49\textwidth]{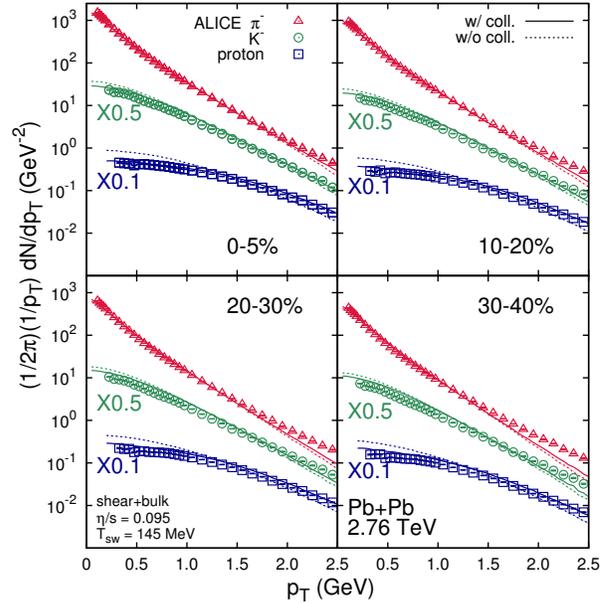}
\caption{
$p_T$ spectra of identified hadrons
for centrality classes $0-5\%$, $10-20\%$, $20-30\%$ and $30-40\%$
of Pb-Pb collisions
with $\sqrt{s_{\scriptsize \textrm{NN} } } = 2.76\,\textrm{TeV}$.
The solid curves and the dashed curves correspond to full \texttt{UrQMD} and
\texttt{UrQMD} without collisions, respectively.
The statistical errors in the calculation are shown
as the bands around the curves.
The ALICE data \cite{Abelev:2013vea} are shown for comparison.
}
\label{dNdpT_pid_casc_LHC}
\end{center}
\end{figure}

\begin{figure}[tbp]
\begin{center}
\includegraphics[width=0.49\textwidth]{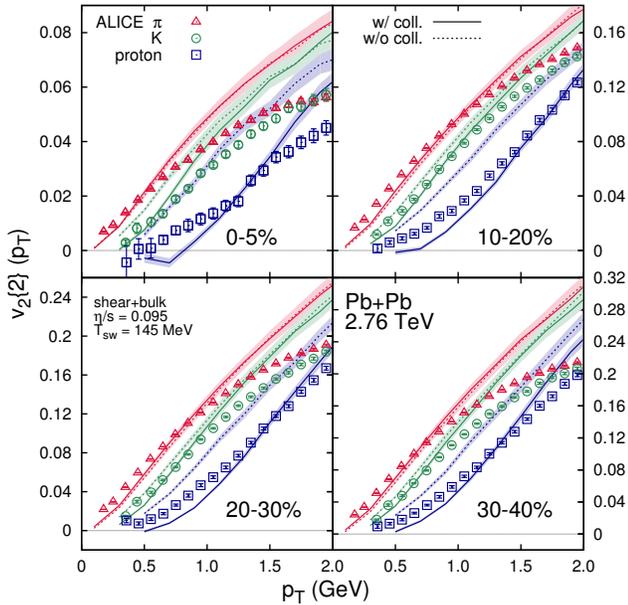}
\caption{
$p_T$-differential $v_2 \{ 2 \}$ of identified hadrons
for centrality classes $0-5\%$, $10-20\%$, $20-30\%$ and $30-40\%$
of Pb-Pb collisions
with $\sqrt{s_{\scriptsize \textrm{NN} } } = 2.76\,\textrm{TeV}$.
The solid curves and the dashed curves correspond to full \texttt{UrQMD} and
\texttt{UrQMD} without collisions, respectively.
The statistical errors in the calculation are shown
as the bands around the curves.
The ALICE data \cite{ALICE:2014v2id} are shown for comparison.
}
\label{v2pT_pid_casc_LHC}
\end{center}
\end{figure}

The $p_T$ differential $v_n\{ 2 \}$ of $n = 2,3,4$ of charged hadrons
are compared with the ALICE \cite{ALICE:2011ab}
and the CMS \cite{CMS:2013v2,CMS:2014vn} data in Fig.~\ref{vnpT_hch_LHC}.
Note that the $p_T$-differential $v_n$ is evaluated
from the azimuthal correlation
between particles of interest and reference flow particles,
given that the particles of interest are those
in specific $p_T$ bins \cite{Bilandzic:2010jr}.
Although $v_2\{ 2 \}$ deviates from the data at high $p_T$,
especially when compared with the ALICE measurements,
our calculation shows a reasonable agreement with data for $p_T\lesssim 1$~GeV,
where we have the most particles.

We next turn to identified hadron observables at LHC energies.
The $p_T$-differential spectra of pions, kaons and protons are shown
in Fig.~\ref{dNdpT_pid_casc_LHC}, with (solid line)
and without (dashed line) the effect of hadronic rescattering,
for four different centralities.  
Calculations that include hadronic rescattering agree very well
with measurements for the most central collisions ($0-5\%$),
for all three hadron species.
Tension with data appears and increases in more peripheral centralities,
especially in kaons and protons, but also in pions at $p_T$ above $1.5-2$~GeV.
As expected from the discussion of integrated observables,
the hadronic transport phase has a minor effect on the pion spectra,
which is slightly hardened at $p_T > 2$ GeV.
The kaon spectra get flatter resulting in a better agreement
with the experimental measurement.
A more significant effect of rescattering is seen in the proton spectra:
the low $p_T$ parts of the spectra is reduced in the transport phase
owing to $B\bar{B}$ annihilations while hadronic rescattering shifts
more protons to higher $p_T$.
This shows once again that the inclusion of the hadronic transport phase is
important to describe the measured proton spectra at the LHC. 

Figure \ref{v2pT_pid_casc_LHC} shows
identified particle elliptic flow coefficients at the LHC,
with measurements from the ALICE collaboration~\cite{ALICE:2014v2id}.
Comparing the simulation results with (solid lines)
and without (dashed lines) hadronic rescattering,
we find once more that pions and kaons $v_2(p_T)$ are
largely insensitive to rescattering.
On the other hand, hadronic rescattering has a large effect
on the proton $v_2 (p_T)$,
which is considerably decreased by hadronic interactions.
Even though $v_2$ around the mean $p_T$ is well reproduced,
our calculations overestimate the $v_2$ of pions and kaons at higher $p_T$.
We highlight that tension with ALICE measurements was also observed
at high $p_T$ for the $v_2$ of charged hadrons
shown in Fig.~\ref{vnpT_hch_LHC}.
We note that tension with measurements at high $p_T$ is less worrying
than in lower regions of transverse momenta,
since this region of $p_T$ is more sensitive to uncertainties
in the viscous corrections to the hadron distribution function ($\delta f$),
as well as potential contribution from recombination
with (mini-)jet shower partons.
Nevertheless, there still seems to be room for improvement
at lower $p_T$ in our description of identified hadron $v_n$.

\begin{figure}[tbp]
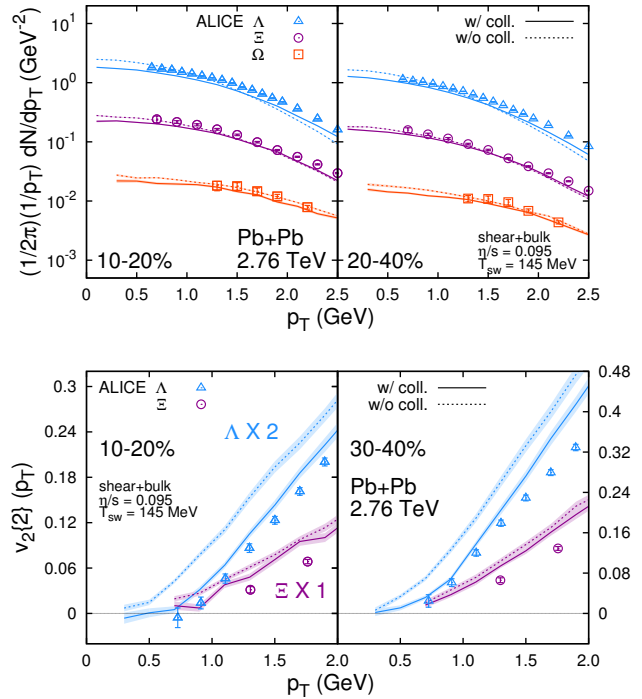

\begin{center}
\includegraphics[width=0.49\textwidth]{dNdpT_str_1040.pdf}\\
\includegraphics[width=0.49\textwidth]{v2pT_str_1040_new1.pdf}
\caption{
$p_T$ spectra (upper) and $p_T$-differential $v_2 \{ 2 \}$ (lower)
of strange baryons of Pb-Pb collisions
with $\sqrt{s_{\scriptsize \textrm{NN} } } = 2.76\,\textrm{TeV}$.
The solid curves and the dashed curves correspond to full \texttt{UrQMD} and
\texttt{UrQMD} without collisions, respectively.
The statistical errors in the calculation are shown
as the bands around the curves.
}
\label{dNdpT_v2pT_str_1040_LHC}
\end{center}
\end{figure}

The $p_T$ spectra and $v_2$ of strange baryons are shown
in Fig.~ \ref{dNdpT_v2pT_str_1040_LHC} and compared
with the ALICE data~\cite{ALICE:2014mstr,Abelev:2013Ks0L,ALICE:2014v2id}.
The $p_T$ dependence of the spectra of $\Lambda$, $\Xi$ and $\Omega$
is described well,
although deviations of up to $20\%$ are observed in the normalization.
The effect of hadronic rescattering,
which suppresses the $p_T$ spectra more at low $p_T$,
is consistent with the decrease in multiplicity
and the increase in average transverse momentum seen in the previous section.
We consider the level of agreement with experimental data to be acceptable
considering the non-negligible dependence of heavier hadrons
on the switching temperature between hydrodynamics and UrQMD
shown previously in Fig.~\ref{dNdy_pTave_vsTtr_0005}.

For all three heavy strange baryons,
our calculation overestimates the $v_2(p_T)$.
Previous studies \cite{Zhu:2015dfa,Takeuchi:2015ana},
also based on a hybrid approach with isothermal particlization,
found some tension with hyperons as well,
although we highlight that comparisons with these previous models
is not straightforward because of differences in the hydrodynamic modeling
(e.g. initial conditions).
Once again, since heavy hadrons have been shown in this work
to be especially sensitive to the transition between hydrodynamics
and the afterburner, this tension is not unexpected.
There have been proposals in the literature that strange hadrons 
may chemically freeze out earlier than non-strange 
particles \cite{vanHecke:1999jh,Arbex:2001vx,Chatterjee:2013yga}.
We cannot necessarily conclude this from our investigations, 
but we can say that improvements in the transition
between hydrodynamics and hadronic transport
are important to obtain a better description of heavy hadrons,
including the $\Lambda$, $\Xi$ and $\Omega$ baryons.

\begin{figure}[tbp]
\begin{center}
\includegraphics[width=0.49\textwidth]{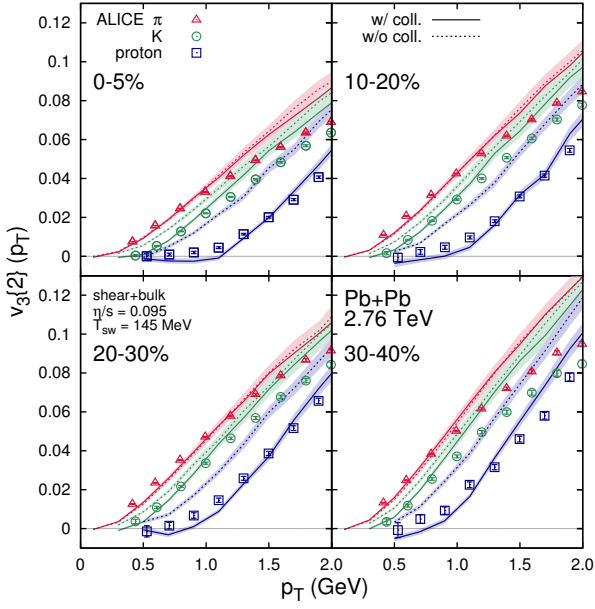}
\caption{
$p_T$-differential $v_3 \{ 2 \}$ of pions, kaons and anti-protons
for centrality classes $0-5\%$, $10-20\%$, $20-30\%$ and $30-40\%$
of Pb-Pb collisions
with $\sqrt{s_{\scriptsize \textrm{NN} } } = 2.76\,\textrm{TeV}$.
The statistical errors in the calculation are shown
as the bands around the curves. Measurements are from \cite{Adam:2016nfo}.
}
\label{v3pT_pid_casc_LHC}
\end{center}
\end{figure}

\begin{figure}[tbp]
\begin{center}
\includegraphics[width=0.49\textwidth]{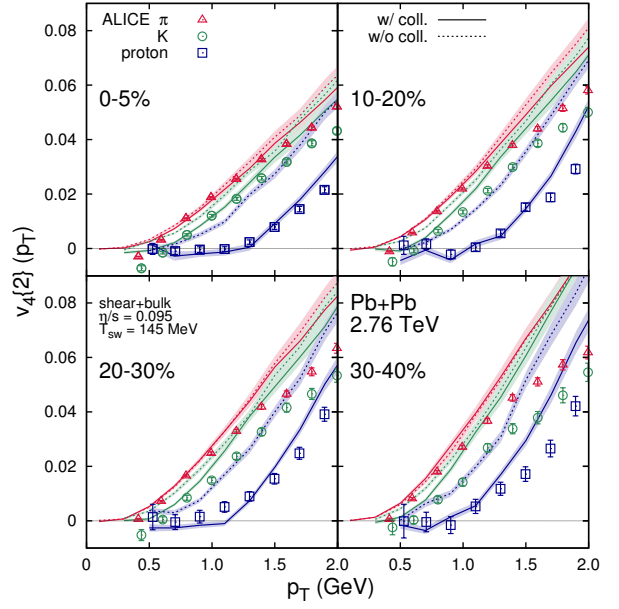}
\caption{
$p_T$-differential $v_4 \{ 2 \}$ of pions, kaons and anti-protons
for centrality classes $0-5\%$, $10-20\%$, $20-30\%$ and $30-40\%$
of Pb-Pb collisions
with $\sqrt{s_{\scriptsize \textrm{NN} } } = 2.76\,\textrm{TeV}$.
The statistical errors in the calculation are shown
as the bands around the curves. Measurements are from \cite{Adam:2016nfo}.
}
\label{v4pT_pid_casc_LHC}
\end{center}
\end{figure}

As shown in Fig.~\ref{v2pT_pid_casc_LHC}
the hadronic rescattering has a significant effect on proton $v_2$ and
we expect a similar effect in the higher harmonics as well.
Figures \ref{v3pT_pid_casc_LHC} and \ref{v4pT_pid_casc_LHC} show
our calculations for the identified hadron $v_3 \{2\}$ and $v_4\{2\}$
in several centrality classes.
Overall, we find that the level of agreement with data and
the effect of rescattering is similar for $v_3$ and $v_4$ to
what was observed for $v_2$.


\begin{figure}[h!]
\begin{center}
\includegraphics[width=0.49\textwidth]{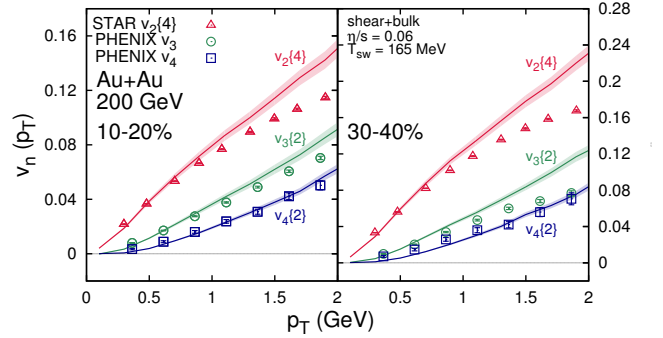}
\caption{
$p_T$-differential $v_2 \{ 4 \}$ and
$v_n \{ 2 \}$ ($n = 3 \,\,\textrm{and} \,\, 4$) of charged hadrons
for centrality classes $10-20\%$ (left) and $30-40\%$ (right)
of Au-Au collisions
with $\sqrt{s_{\scriptsize \textrm{NN} } } = 200\,\textrm{GeV}$.
The statistical errors in the calculation are shown
as the bands around the curves.
The PHENIX \cite{PHENIX:2011vnch} and STAR \cite{STAR:2008vn} data are shown
for comparison.
}
\label{vnpT_hch_RHIC}
\end{center}
\end{figure}

\begin{figure}[h!]
\begin{center}
\includegraphics[width=0.49\textwidth]{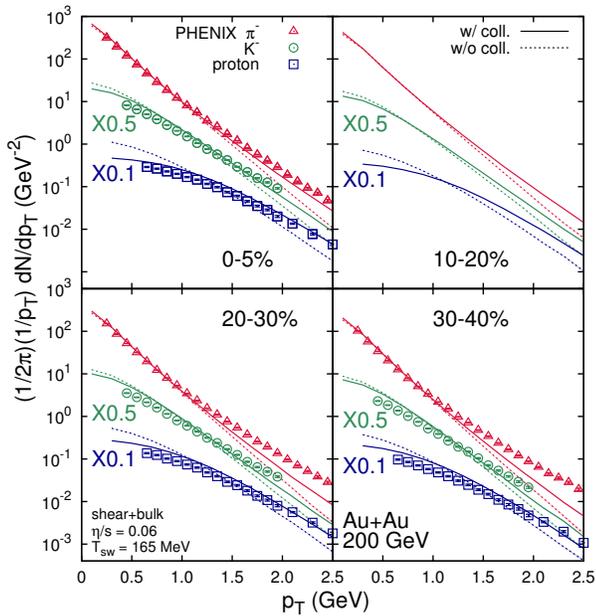}
\caption{
$p_T$ spectra of identified hadrons
for centrality classes $0-5\%$, $10-20\%$, $20-30\%$ and $30-40\%$
of Au-Au collisions
with $\sqrt{s_{\scriptsize \textrm{NN} } } = 200\,\textrm{GeV}$.
The solid curves and the dashed curves correspond to full \texttt{UrQMD} and
\texttt{UrQMD} without collisions, respectively.
The statistical errors in the calculation are shown
as the bands around the curves.
The PHENIX \cite{PHENIX:2004idsp} data are shown for comparison.
}
\label{dNdpT_pid_RHIC}
\end{center}
\end{figure}
\begin{figure}[h!]
\begin{center}
\includegraphics[width=0.49\textwidth]{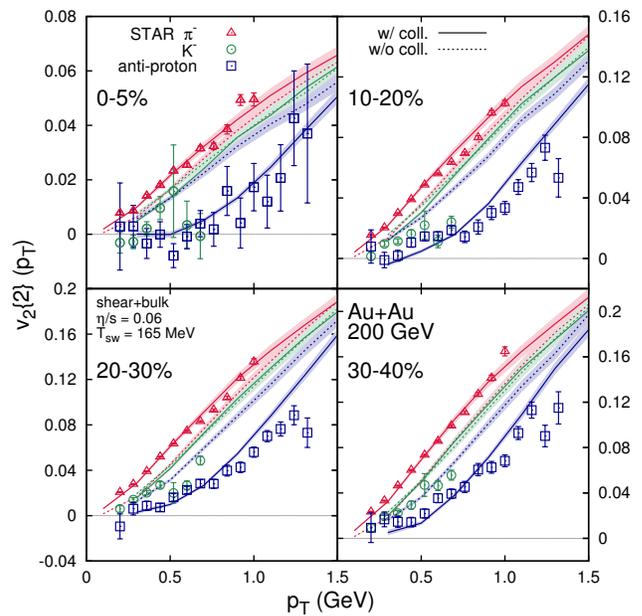}
\caption{
$p_T$-differential $v_2 \{ 2 \}$ (right panels) of identified hadrons
for centrality classes $0-5\%$, $10-20\%$, $20-30\%$ and $30-40\%$
of Au-Au collisions
with $\sqrt{s_{\scriptsize \textrm{NN} } } = 200\,\textrm{GeV}$.
The solid curves and the dashed curves correspond to full \texttt{UrQMD} and
 \texttt{UrQMD} without collisions, respectively.
The statistical errors in the calculation are shown
as the bands around the curves.
The STAR \cite{Adams:2004bi} data are shown for comparison.
}
\label{v2pT_pid_RHIC}
\end{center}
\end{figure}
\begin{figure}[h!]
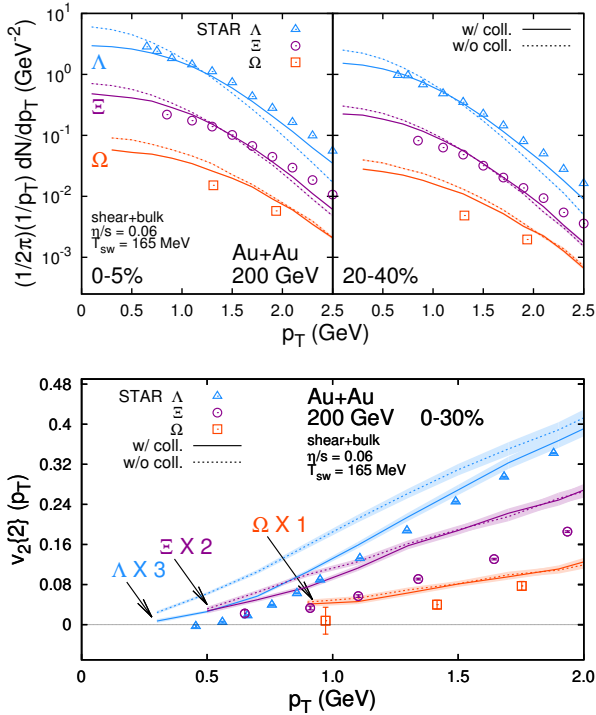

\begin{center}
\includegraphics[width=0.49\textwidth]{dNdpT_str_twoCent.pdf}\\
\includegraphics[width=0.49\textwidth]{v2pT_str_0030C_new1.pdf}
\caption{
$p_T$ spectra (upper) and $p_T$-differential $v_2 \{ 2 \}$ (lower)
of strange baryons in Au-Au collisions
with $\sqrt{s_{\scriptsize \textrm{NN} } } = 200\,\textrm{GeV}$.
The solid curves and the dashed curves correspond to full \texttt{UrQMD} and
\texttt{UrQMD} without collisions, respectively.
The statistical errors in the calculation are shown
as the bands around the curves.
The STAR data on $p_T$ spectra \cite{STAR:2007st}
and $v_2 (p_T)$ \cite{STAR:2016vs}
are shown for comparison.
}
\label{dNdpT_v2pT_str_RHIC}
\end{center}
\end{figure}

We now turn our attention to  
Au-Au collisions
with $\sqrt{s_{\scriptsize \textrm{NN} } } = 200\,\textrm{GeV}$ at RHIC.
Figure \ref{vnpT_hch_RHIC} shows the $p_T$-differential $v_2$, $v_3$, and $v_4$
of charged hadrons.
The $p_T$ spectra and differential $v_2$ of identified hadrons
are shown in Figs.~\ref{dNdpT_pid_RHIC} and \ref{v2pT_pid_RHIC}, respectively.
The hybrid approach provides a good description
of the charged hadron $v_n$ at $p_T\lesssim 1$~GeV.
At higher $p_T$, $v_2\{4\}$ and $v_3\{2\}$ measurements are overestimated.
This is a similar trend as seen at the LHC in Fig.~\ref{vnpT_hch_LHC}.
The agreement observed with data
for the identified spectra (Fig.~\ref{dNdpT_pid_RHIC}) is also comparable
with the LHC results:
calculations describe well the measurements in central collisions
but increasing tension is seen in more peripheral bins.
As for the identified hadron $v_2$ (Fig.~\ref{v2pT_pid_RHIC}),
we highlight that agreement is distinctly better at RHIC than at the LHC,
especially for pions.

A larger tension with experimental data is found for the $p_T$ spectra
and $p_T$-differential elliptic flow of strange baryons,
as shown in Fig.~\ref{dNdpT_v2pT_str_RHIC}.
We repeat that these heavier hadrons are more sensitive to the transition
between hydrodynamics and UrQMD than lighter ones,
and that this level of agreement with measurements is not unexpected.

\section{Summary and conclusion}
\label{sec:summary}

In this paper, we compared a hybrid model of IP-Glasma initial conditions,
shear and bulk viscous hydrodynamics (\texttt{MUSIC}),
and microscopic hadronic transport (\texttt{UrQMD})
with a wide range of integrated and differential measurements
from Pb-Pb collisions
($\sqrt{s_{NN}} = 2.76\,\textrm{TeV}$) at the LHC
and Au-Au collisions ($\sqrt{s_{NN}} = 200\,\textrm{GeV}$) at RHIC. 
We investigated how different observables depend on our model parameters,
such as the transport coefficients and the switching temperature
from hydrodynamics to the hadronic transport.
We found that the bulk viscosity is important to
consistently describe the mid-rapidity multiplicity,
mean $p_T$ of identified hadrons, and the integrated $v_n$ within this model.

The inclusion of the bulk viscosity reduces our estimate
of the value of the effective shear viscosity by approximately $50\%$.
This reduction of shear viscosity is consistent with the intuition that
both the shear and bulk viscosities act to reduce the anisotropic flow,
and that to produce a similar amount of entropy generated
by the larger shear viscosity alone,
the shear viscosity in the presence of non-zero bulk viscosity
should be smaller.

Heavy hadrons were found to be particularly sensitive
to the switching temperature between hydrodynamics and the afterburner.
Future improvements on the matching between hydrodynamics
and the hadronic transport will be important in reducing this dependence.

It should be emphasized that all three components of our model,
IP-Glasma, viscous
hydrodynamics and the hadronic after-burner play important roles.
The energy deposition mechanism of the IP-Glasma model helps
provide a good description of the correct higher flow harmonics,
and the large gradient
found in the initial energy density enhance the effect of bulk viscosity. 
The hadronic afterburner is important
to improve the description of identified particle observables.

Looking ahead, the addition of  mini-jets and jet energy loss will allow us to
extend the investigations presented in this work in the intermediate and 
high $p_T$ regions of the observables.
Moreover in addition to the observables described above,
it is also useful to study the effects of fluctuations
and transport coefficients
on the event plane correlations and flow harmonics correlations $r_n$,
which is the subject of a future publication.
A further area of possible 
improvement is the treatment of the non-equilibrium corrections to
the thermal distribution functions.
So far, our $\delta f$ is species-independent.
Making it species-dependent following the line of arguments
in Ref.~\cite{Molnar:2014fva}, for example,
is an undertaking we leave for future work.

\section*{Acknowledgments}
This work was supported in part
by the Natural Sciences and Engineering Research Council of Canada.
SR acknowledges funding of a Helmholtz Young Investigator Group VH-NG-822
from the Helmholtz Association and GSI.
JFP was supported in part by the U.S. D.O.E. Office of Science,
under Award No. DE-FG02-88ER40388. 
BPS was supported under DOE contract  No. DE-SC0012704
and acknowledges a DOE Office of Science Early Career Award.
CG gratefully acknowledges support from the Canada Council for the Arts
through its Killam Research Fellowship program.
Computations were performed on the Guillimin supercomputer
at McGill University under the auspices of
Calcul Quebec and Compute Canada.
The operation of Guillimin is funded by the Canada Foundation for Innovation
(CFI), the National Science and Engineering Research Council (NSERC),
NanoQuebec, and the Fonds Quebecois
de Recherche sur la Nature et les Technologies (FQRNT).
This research used resources of
the National Energy Research Scientific Computing Center,
which is supported by the Office of Science of the U.S. Department of Energy
under Contract No. DE-AC02- 05CH11231.

\appendix

\section{Viscous corrections to the momentum distribution of hadrons}

\label{sec:appendixdf}

At the end of a hydrodynamic simulation,
fluid elements must be converted into hadronic degrees of freedom.
This conversion is made possible under the assumption
that hydrodynamics and kinetic theory have an overlapping region of validity
in the late stage of the collision.
This overlap allows for the momentum distribution of hadrons to be related
to the energy-momentum tensor of the fluid in such a way
that energy and momentum are conserved across this transition.

From a kinetic theory point of view,
the energy-momentum tensor $T^{\mu\nu}$ only contains information
about the second moment of the momentum distribution function,
which constrains predominantly the small momentum region of the distribution.
In consequence, the transition from fluid to particles carries some ambiguity,
since multiple hadronic momentum distributions
similar at lower momentum but different at higher momentum
can correspond to the same energy-momentum tensor.
On the other hand this uncertainty in the higher momentum region
of the distribution should not be a major issue for soft hadronic observables,
which are the observables of interest
in hydrodynamic simulations of heavy ion collisions.
 
The matching from fluid to hadrons also depends on the collision kernel
describing the microscopic interactions of all species of hadrons,
which is not known well.
Even simplified description of species dependence of hadronic interactions
can become quite challenging to handle (see e.g. Ref.~\cite{Molnar:2014fva}).
In consequence, simpler approximations are generally made
regarding the collision kernel
describing the microscopic interactions of hadrons
in order to relate the energy-momentum tensor
to the hadron's momentum distribution.
In this work, the relaxation time approximation
and the 14-moments approximation were both used to this effect.

An additional assumption made
regarding the dependence of the hadronic momentum distribution
on the shear stress tensor $ \pi^{\mu\nu}$ and the bulk pressure $\Pi$ is
that it can be linearized:
\begin{eqnarray}
	f(P,\pi^{\mu\nu},\Pi) & \approx &
		f^{(0)}(P) + \mathcal{C}_{\textrm{shear}}(P) \pi^{\mu\nu} P_\mu P_\nu
		\nonumber \\
		& & +\mathcal{C}_{\textrm{bulk}}(P) \Pi \nonumber \\
		& \approx & f^{(0)}(P) + \delta f_{\textrm{shear}}
		+ \delta f_{\textrm{bulk}}
\end{eqnarray}
where we used the common notation that the linearized term
depending on the shear stress tensor $\pi^{\mu\nu}$ is referred to
as $\delta f_{\scriptsize \textrm{shear}}$ and
the one depending on the bulk pressure $\Pi$ is
$\delta f_{\scriptsize \textrm{bulk}}$.
The functional form of $\mathcal{C}_{\scriptsize \textrm{shear}}(P)$
and $\mathcal{C}_{\scriptsize \textrm{bulk}}(P)$ depend
on the collision kernel used to describe hadronic interactions.
The explicit form of $\delta f_{\scriptsize \textrm{shear/bulk}}$ is given
by Eqs~(\ref{eq:dfshear}) and (\ref{eq:dfbulk}).

Since there is a certain level of uncertainty
in the determination of $\delta f_{\scriptsize \textrm{shear}}$
and $\delta f_{\scriptsize \textrm{bulk}}$ from the energy-momentum tensor,
it is useful to quantify the dependence of hadronic observables
on these two quantities.
In this section, this is done for the integrated observables
shown in Section~\ref{sec:integrated} for the LHC.
Since the effect of the bulk pressure and the shear stress tensor
on integrated hadronic observables are significantly different,
they are discussed separately in this Appendix.

\subsection{Corrections from shear viscosity}

\begin{figure}[t]
\begin{center}
\includegraphics[width=0.37\textwidth]{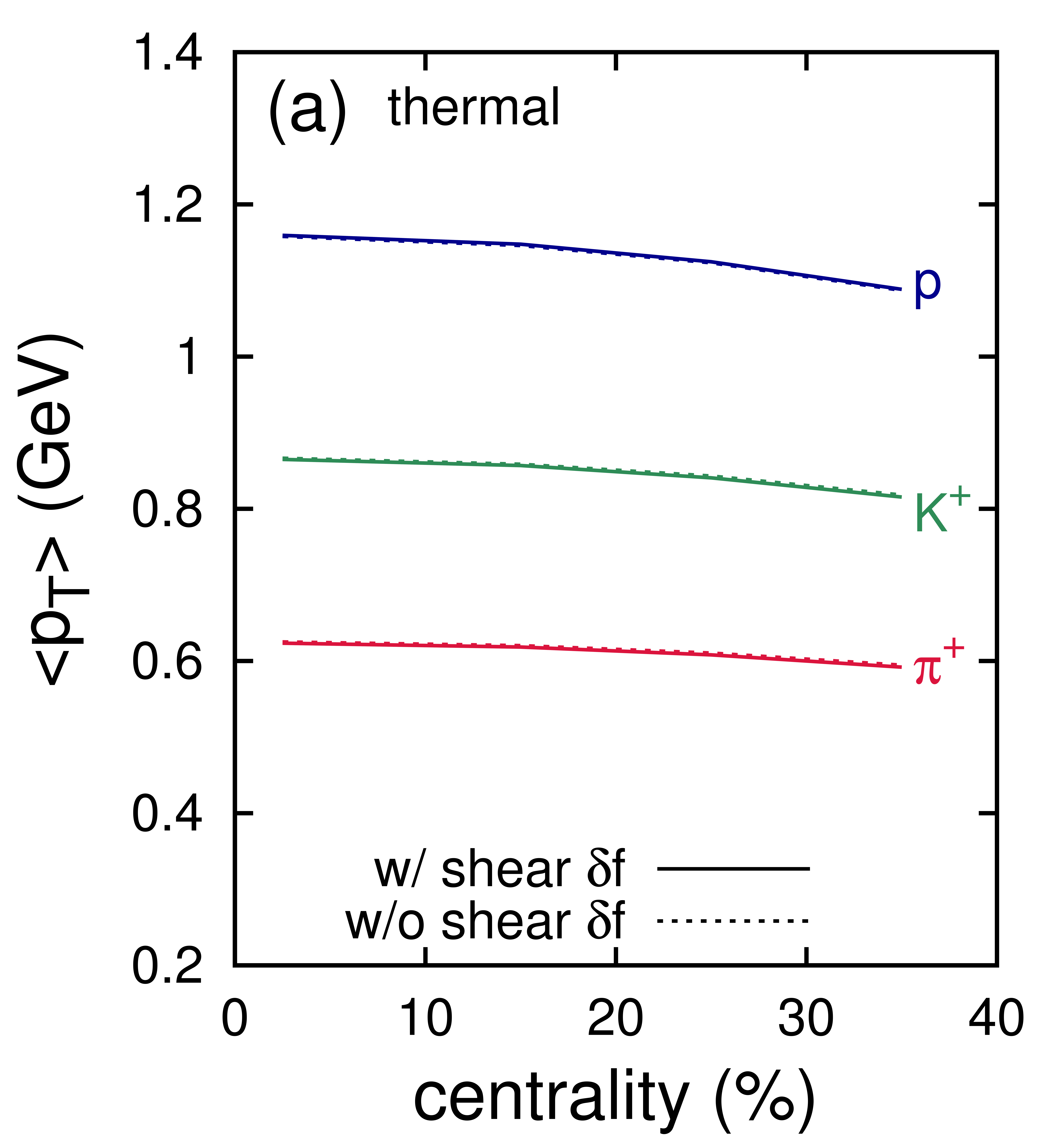}
\includegraphics[width=0.37\textwidth]{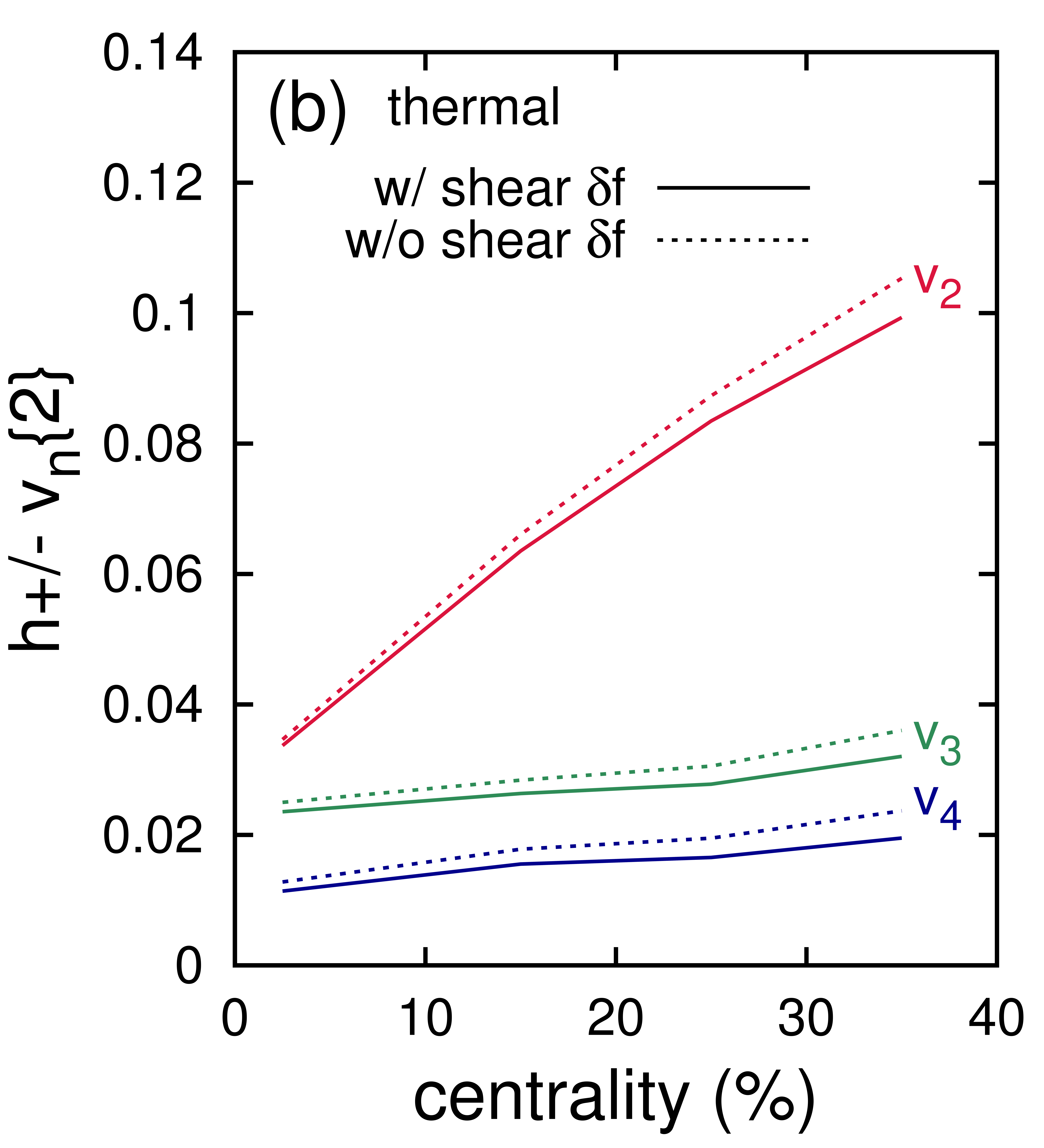}
\caption{
Effect of $\delta f_{\textrm{shear}}$ on
(a) the average transverse momentum of thermally emitted (Cooper-Frye) pions,
kaons and protons, and
(b) the $p_T$-integrated $v_2$, $v_3$ and $v_4$ or charged hadrons,
as a function of centrality, for Pb-Pb collisions at $\sqrt{s_{NN}}=2.76$~TeV.
}
\label{fig:sheardf}
\end{center}
\end{figure}

Because of the tensor structure
of the shear $\pi^{\mu\nu}$-linearized momentum distribution
the multiplicity $dN/dy$ does not depend
on $\delta f_{\scriptsize \textrm{shear}}$ (for a boost-invariant system).
Experiments often employ cuts in transverse momentum
when calculating the multiplicity of hadrons,
which will lead to some dependence on $\delta f_{\scriptsize \textrm{shear}}$.
Nevertheless, these cuts are very small for the measurements used in this work
and we verified numerically that the multiplicity of hadrons is
essentially identical with and without $\delta f_{\scriptsize \textrm{shear}}$.

The effect of $\delta f_{\scriptsize \textrm{shear}}$
on the average transverse momentum of thermal (Cooper-Frye) pions,
kaons and protons and the $p_T$-integrated $v_n$ of thermal charged hadrons
is shown in Figs.~\ref{fig:sheardf}(a) and (b) respectively.
The effect on $\langle p_T \rangle$ is very small,
and we verified that it remains small
even after hadronic decays are taken into account.
The $v_n$ of charged hadrons displays a larger dependence
on $\delta f_{\scriptsize \textrm{shear}}$, of $2-5\%$ for the $v_2$, $5-10\%$
for the $v_3$ and $10-20\%$ for the $v_4$.
We verified that we obtain similar numbers after hadronic decays are included.

\begin{figure}[t]
\begin{center}
\includegraphics[width=0.37\textwidth]{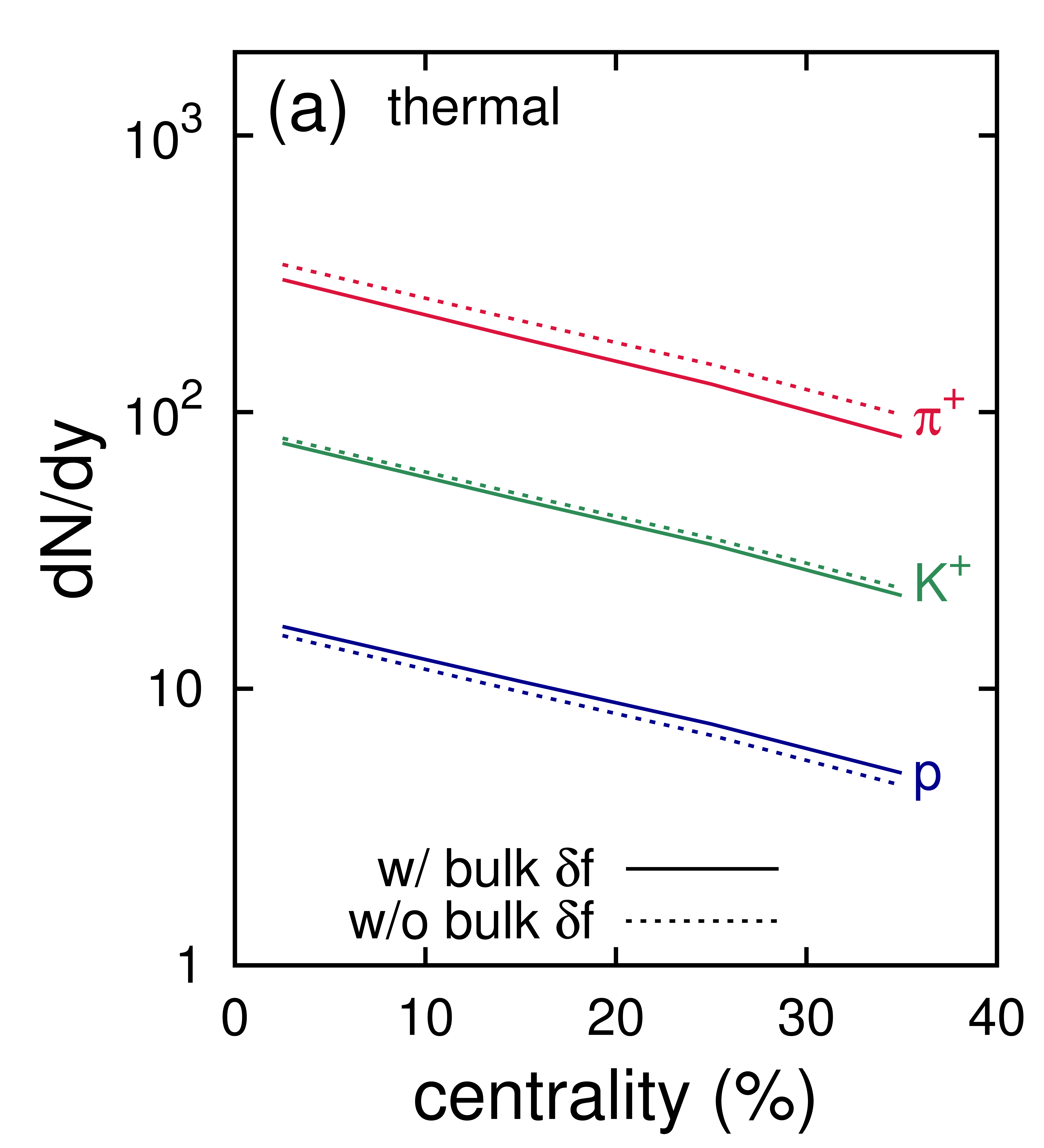}
\includegraphics[width=0.37\textwidth]{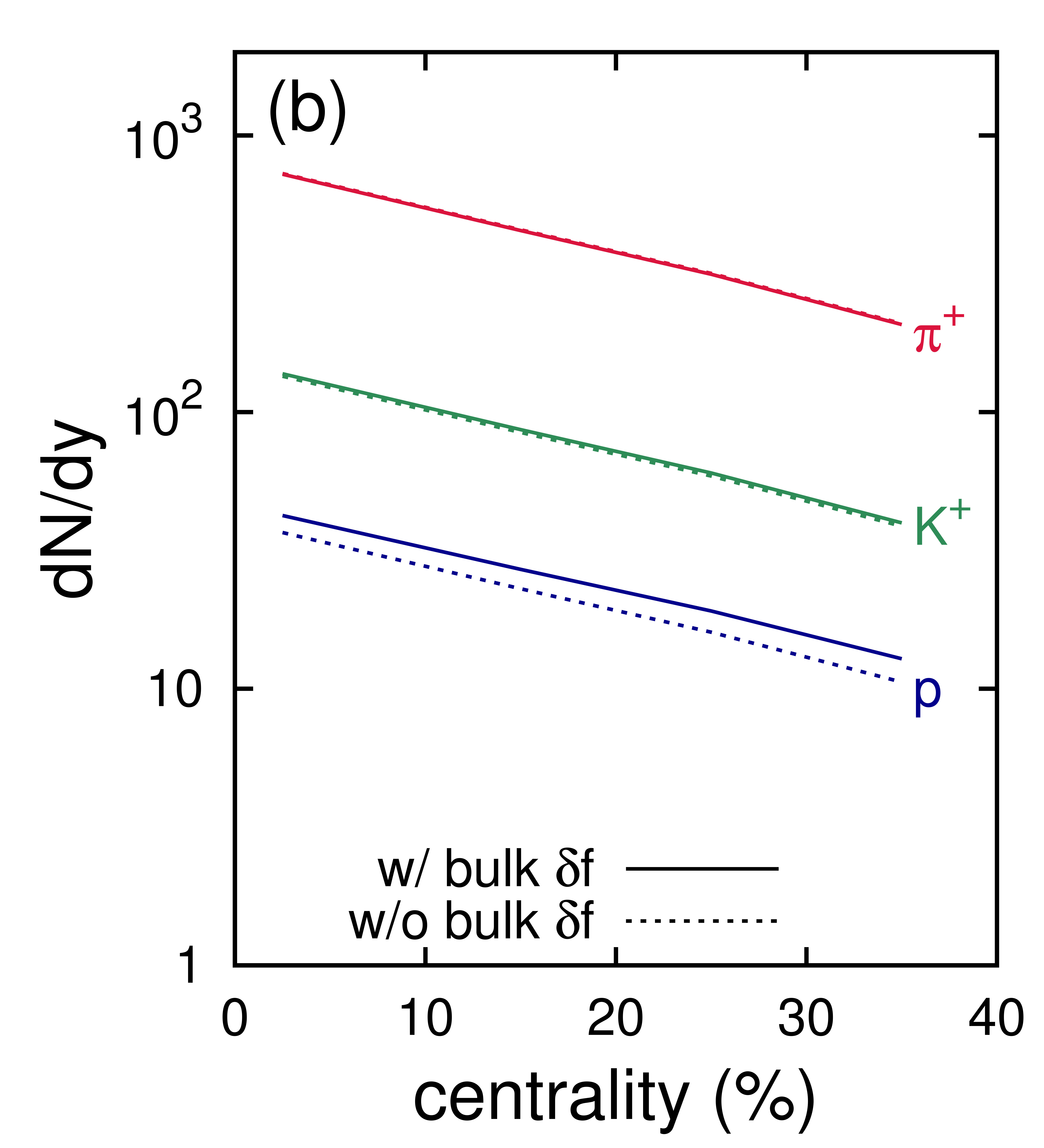}\\
\caption{
Effect of $\delta f_{\textrm{bulk}}$ on the pion, kaon and proton $dN/dy$
as a function of centrality, for Pb-Pb collisions at $\sqrt{s_{NN}}=2.76$~TeV.
(a) is for  thermal (Cooper-Frye) hadrons, (b) is after hadronic decays.
}
\label{fig:bulk_df_mult}
\end{center}
\end{figure}

\begin{figure}[t]
\begin{center}
\includegraphics[width=0.37\textwidth]{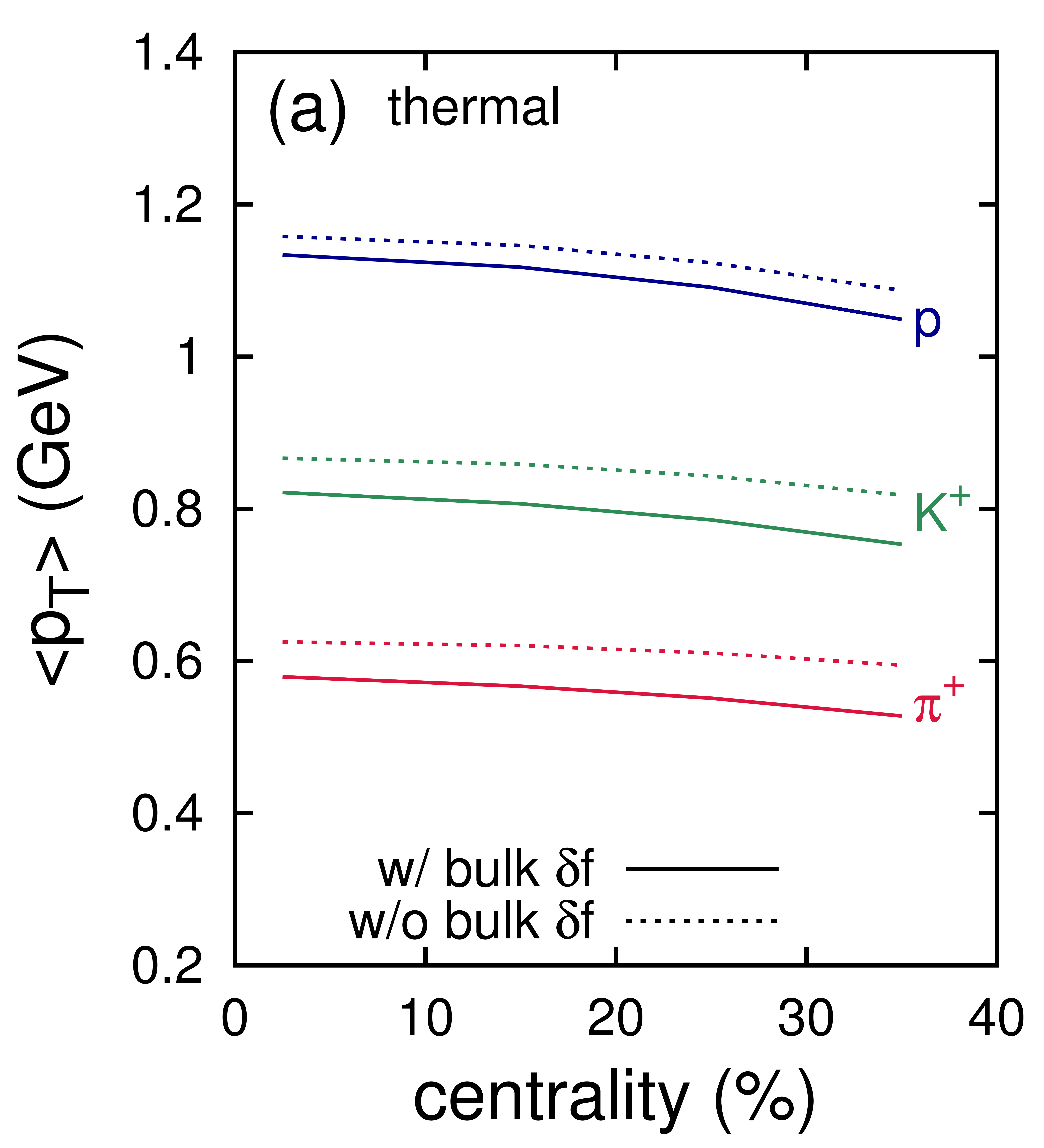}
\includegraphics[width=0.37\textwidth]{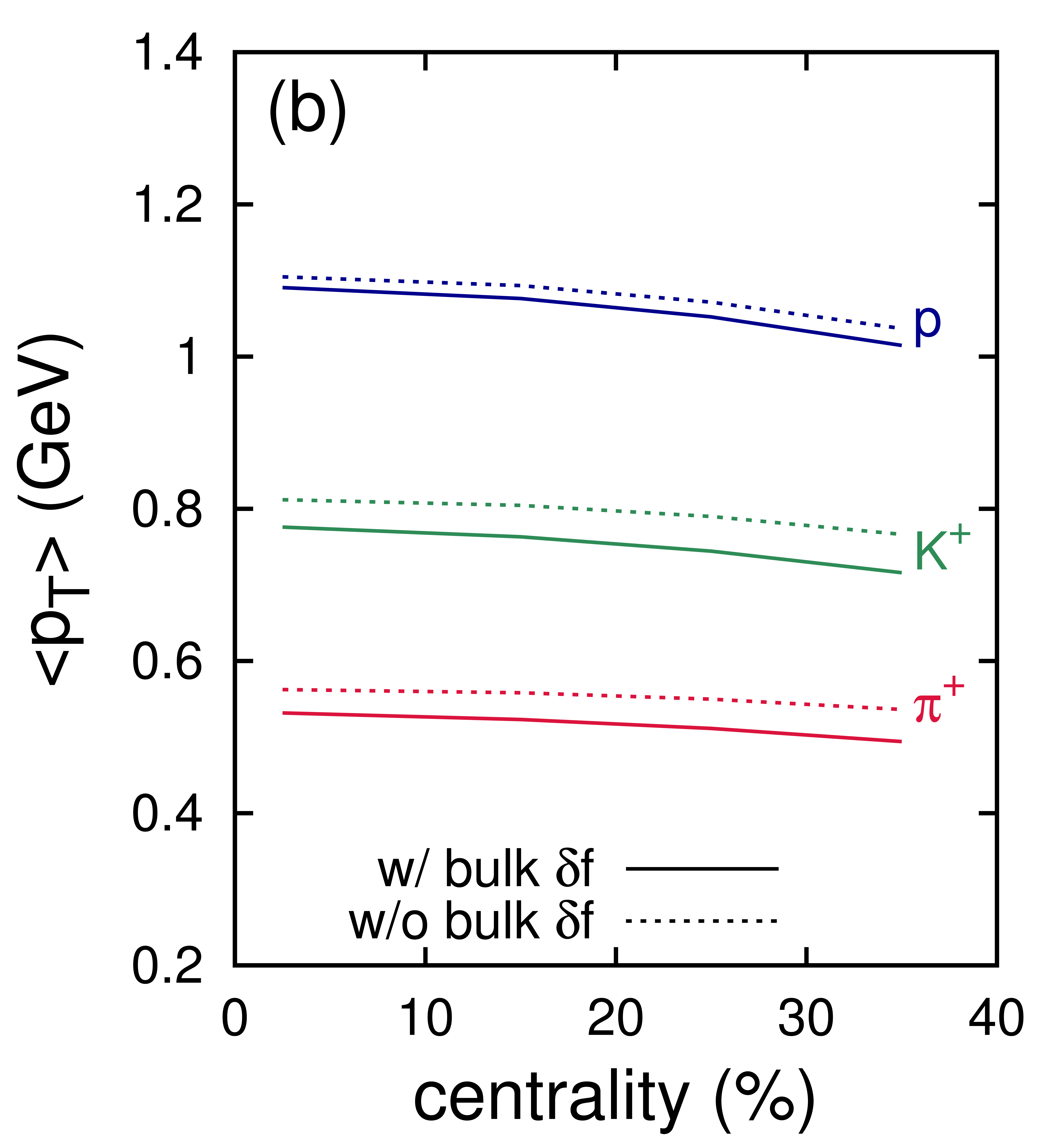}\\
\caption{
Effect of $\delta f_{\textrm{bulk}}$ on the average transverse momentum
of pions, kaons and protons as a function of centrality,
for Pb-Pb collisions at $\sqrt{s_{NN}}=2.76$~TeV.
(a) is for  thermal (Cooper-Frye) hadrons, (b) is after hadronic decays.
}
\label{fig:bulk_df_mean}
\end{center}
\end{figure}

\subsection{Corrections from bulk viscosity}

The $\delta f_{\scriptsize \textrm{bulk}}$ used in this work
has an explicit dependence on the mass of hadrons --- see Eq.(\ref{eq:dfbulk}).
Consequently, it is to be expected
that the effect of $\delta f_{\scriptsize \textrm{bulk}}$
on different species of hadron will show a mass dependence.

The effect of $\delta f_{\scriptsize \textrm{bulk}}$
on the pion, kaon and proton $dN/dy$ is shown in Fig.~\ref{fig:bulk_df_mult}.
Figure~\ref{fig:bulk_df_mult}(a) is for thermal hadrons,
and (b) is after hadronic decays.
The effect of $\delta f_{\scriptsize \textrm{bulk}}$ decreases
the multiplicity of thermal pions by $\sim 15\%$,
very slightly decreases the multiplicity of thermal kaons ($\sim 5\%$),
and increases the multiplicity of protons by $\sim 10\%$.
There is thus a change in the effect of $\delta f_{\scriptsize \textrm{bulk}}$
on the multiplicity at a mass slightly above the kaon mass.
Hadronic decays cancel out the suppression
from $\delta f_{\scriptsize \textrm{bulk}}$ on thermal pions
against the enhancement from $\delta f_{\scriptsize \textrm{bulk}}$
on heavier hadrons which decay into pions.
The result is a negligible effect of $\delta f_{\scriptsize \textrm{bulk}}$
on the final pion multiplicity.
A similar effect is seen in the final kaon multiplicity,
while the already-enhanced thermal proton multiplicity is further increased
after decays ($\sim 20\%$)
by the effect of $\delta f_{\scriptsize \textrm{bulk}}$ on heavier hadrons.

The effect of $\delta f_{\scriptsize \textrm{bulk}}$
on the average transverse momentum of pions, kaons and protons
also shows a mass dependence,
as seen in Fig.~\ref{fig:bulk_df_mean}, with
(a) being once again for thermal hadrons and
(b) being the result including hadronic decays.
The thermal pion $\langle p_T \rangle$ is suppressed by $\sim 10\%$,
while thermal protons are suppressed by $\sim 5\%$ and kaons are in-between.
Since heavier hadrons have a smaller correction
from $\delta f_{\scriptsize \textrm{bulk}}$,
the inclusion of hadronic decays lessens the effect of
$\delta f_{\scriptsize \textrm{bulk}}$ on lighter hadrons,
as seen in Fig.~\ref{fig:bulk_df_mean}(b).

Finally, unlike for $\delta f_{\scriptsize \textrm{shear}}$,
we found that the $\delta f_{\scriptsize \textrm{bulk}}$ used in this work
leaves the $p_T$-integrated $v_n$ of charged hadrons unchanged,
whether hadronic decays are included or not. 

We highlight that the multiplicity and average transverse momentum of hadrons,
which were largely insensitive to $\delta f_{\scriptsize \textrm{shear}}$,
are affected by $\delta f_{\scriptsize \textrm{bulk}}$.
We thus find the interesting conclusion that integrated observables
that are not sensitive to $\delta f_{\scriptsize \textrm{shear}}$ are sensitive
to $\delta f_{\scriptsize \textrm{bulk}}$, and vice versa.
While the effect of $\delta f_{\scriptsize \textrm{bulk}}$ is not very large,
the results found in Figs.~\ref{fig:bulk_df_mult} and~\ref{fig:bulk_df_mean}
certainly warrants additional investigations in the future
about the effect of $\delta f_{\scriptsize \textrm{bulk}}$
on hadronic observables.


\begin{thebibliography}{99}

\bibitem{gale:2013}
  C. Gale, S. Jeon, and B. Schenke, 
  Int.\ J.\ Mod.\ Phys.\ A \textbf{28}, 1340011 (2013). 

\bibitem{deSouza:2015ena} 
  R.~Derradi de Souza, T.~Koide and T.~Kodama,
  Prog.\ Part.\ Nucl.\ Phys.\  {\bf 86}, 35 (2016)


\bibitem{Gyulassy:2004zy} M.~Gyulassy and L.~McLerran, 
  Nucl.\ Phys.\ A \textbf{750}, 30 (2005). 

\bibitem {whitepaper1} BRAHMS collaboration, Nuclear Physics A
\textbf{757}, Issues 1-2, Pages 1-27 (2005).

\bibitem {whitepaper2} PHENIX collaboration, Nuclear Physics A
\textbf{757}, Issues 1-2, Pages 184-283 (2005).

\bibitem {whitepaper3} PHOBOS collaboration, Nuclear Physics A
\textbf{757}, Issues 1-2, Pages 28-101 (2005).

\bibitem {whitepaper4} STAR collaboration,
  Nuclear Physics A \textbf{757}, Issues 1-2, Pages 102-183 (2005).

\bibitem{Romatschke:2007mq} 
  P.~Romatschke and U.~Romatschke,
  Phys.\ Rev.\ Lett.\  {\bf 99}, 172301 (2007).


\bibitem{MUSIC} B.~Schenke, S.~Jeon and C.~Gale, 
Phys.\ Rev.\ C \textbf{82}, 014903 (2010);
Phys.\ Rev.\ Lett.\ \textbf{106}, 042301 (2011); 
Phys.\ Rev.\ C \textbf{85}, 024901 (2012). 

\bibitem{Song:2010aq} 
  H.~Song, S.~A.~Bass and U.~Heinz,
  Phys.\ Rev.\ C {\bf 83}, 024912 (2011)
  doi:10.1103/PhysRevC.83.024912
  [arXiv:1012.0555 [nucl-th]].

\bibitem{Song:2011qa} 
  H.~Song, S.~A.~Bass and U.~Heinz,
  Phys.\ Rev.\ C {\bf 83}, 054912 (2011)
  Erratum: [Phys.\ Rev.\ C {\bf 87}, no. 1, 019902 (2013)]
  doi:10.1103/PhysRevC.83.054912, 10.1103/PhysRevC.87.019902
  [arXiv:1103.2380 [nucl-th]].

\bibitem{Arnold:2006bv}
  P.~Arnold, C.~Dogan, and G.~D.~Moore,
  Phys.\ Rev. D {\bf 74}, 085021 (2006).


\bibitem{Meyer:2007dy} 
  H.~B.~Meyer,
  Phys.\ Rev.\ Lett.\  {\bf 100}, 162001 (2008)
  doi:10.1103/PhysRevLett.100.162001
  [arXiv:0710.3717 [hep-lat]].

\bibitem{Karsch:2007jc} 
  F.~Karsch, D.~Kharzeev and K.~Tuchin,
  Phys.\ Lett.\ B {\bf 663}, 217 (2008).

\bibitem{Buchel:2007mf} 
  A.~Buchel,
  Phys.\ Lett.\ B {\bf 663}, 286 (2008)
  doi:10.1016/j.physletb.2008.03.069
  [arXiv:0708.3459 [hep-th]].

\bibitem{Kharzeev:2008tra}
  D.~Kharzeev and K.~Tuchin,
  JHEP.\ {\bf 09}, 093 (2008).


\bibitem{NoronhaHostler:2008ju} 
  J.~Noronha-Hostler, J.~Noronha and C.~Greiner,
  Phys.\ Rev.\ Lett.\  {\bf 103}, 172302 (2009).

\bibitem{Torrieri:2008ip} 
  G.~Torrieri and I.~Mishustin,
  Phys.\ Rev.\ C {\bf 78}, 021901 (2008)
  [arXiv:0805.0442 [hep-ph]].

\bibitem{Rajagopal:2009yw} 
  K.~Rajagopal and N.~Tripuraneni,
  JHEP {\bf 1003}, 018 (2010)
  [arXiv:0908.1785 [hep-ph]].

\bibitem{Habich:2014tpa} 
  M.~Habich and P.~Romatschke,
  JHEP {\bf 1412}, 054 (2014)
  doi:10.1007/JHEP12(2014)054
  [arXiv:1405.1978 [hep-ph]].

\bibitem{Denicol:2015bpa} 
  G.~S.~Denicol, C.~Gale and S.~Jeon,
  PoS CPOD {\bf 2014}, 033 (2015)
  [arXiv:1503.00531 [nucl-th]].

\bibitem{Denicol:2009am} 
  G.~S.~Denicol, T.~Kodama, T.~Koide and P.~Mota,
  Phys.\ Rev.\ C {\bf 80}, 064901 (2009).

\bibitem{Bozek:2009dw} 
  P.~Bozek,
  Phys.\ Rev.\ C {\bf 81}, 034909 (2010).

\bibitem{Bozek:2012qs} 
  P.~Bozek and I.~Wyskiel-Piekarska,
  Phys.\ Rev.\ C {\bf 85}, 064915 (2012).


\bibitem{Song:2009rh} 
  H.~Song and U.~W.~Heinz,
  Phys.\ Rev.\ C {\bf 81}, 024905 (2010).

\bibitem{Huovinen:2012is} 
  P.~Huovinen and H.~Petersen,
  Eur.\ Phys.\ J.\ A {\bf 48}, 171 (2012)
  doi:10.1140/epja/i2012-12171-9
  [arXiv:1206.3371 [nucl-th]].

\bibitem{Monnai:2009ad} 
  A.~Monnai and T.~Hirano,
  Phys.\ Rev.\ C {\bf 80}, 054906 (2009).

\bibitem{Dusling:2011fd} 
  K.~Dusling and T.~Sch\"afer,
  Phys.\ Rev.\ C {\bf 85}, 044909 (2012).

\bibitem{Noronha-Hostler:2013gga} 
  J.~Noronha-Hostler, G.~S.~Denicol, J.~Noronha, R.~P.~G.~Andrade and
F.~Grassi,
  Phys.\ Rev.\ C {\bf 88}, 044916 (2013).

\bibitem{Noronha-Hostler:2014dqa} 
  J.~Noronha-Hostler, J.~Noronha and F.~Grassi,
  Phys.\ Rev.\ C {\bf 90}, no. 3, 034907 (2014).

\bibitem{Ryu:2015vwa} 
  S.~Ryu, J.-F.~Paquet, C.~Shen, G.~S.~Denicol, B.~Schenke,
  S.~Jeon and C.~Gale,
  Phys.\ Rev.\ Lett.\  {\bf 115}, no. 13, 132301 (2015)
  doi:10.1103/PhysRevLett.115.132301
  [arXiv:1502.01675 [nucl-th]].

\bibitem{Schenke:2012wb} 
  B.~Schenke, P.~Tribedy and R.~Venugopalan,
  Phys.\ Rev.\ Lett.\  {\bf 108}, 252301 (2012).

\bibitem{Denicol:2014vaa} 
  G.~S.~Denicol, S.~Jeon and C.~Gale,
  Phys.\ Rev.\ C {\bf 90}, no. 2, 024912 (2014).


\bibitem{UrQMD1} 
  S.~A.~Bass, M.~Belkacem, M.~Bleicher, M.~Brandstetter, L.~Bravina,
C.~Ernst, L.~Gerland and M.~Hofmann {\it et al.},
  Prog.\ Part.\ Nucl.\ Phys.\  {\bf 41}, 255 (1998).

\bibitem{Bleicher:1999xi} 
  M.~Bleicher, E.~Zabrodin, C.~Spieles, S.~A.~Bass, C.~Ernst, S.~Soff,
L.~Bravina and M.~Belkacem {\it et al.},
  J.\ Phys.\ G {\bf 25}, 1859 (1999).

\bibitem{Denicol:2015nhu} 
  G.~Denicol, A.~Monnai and B.~Schenke,
  Phys.\ Rev.\ Lett.\  {\bf 116}, no. 21, 212301 (2016)
  doi:10.1103/PhysRevLett.116.212301
  [arXiv:1512.01538 [nucl-th]].

\bibitem{Bernhard:2016tnd} 
  J.~E.~Bernhard, J.~S.~Moreland, S.~A.~Bass, J.~Liu and U.~Heinz,
  Phys.\ Rev.\ C {\bf 94}, no. 2, 024907 (2016)
  doi:10.1103/PhysRevC.94.024907
  [arXiv:1605.03954 [nucl-th]].



\bibitem{Bass:2000hy}
  S.~A.~Bass and A.~Dumitru,
  Phys.\ Rev.\ C {\bf 61}, 064909 (2000).

\bibitem{Teaney:2001fl}
  D.~Teaney, J.~Lauret, and E.~Shuryak,
  Phys.\ Rev.\ Lett. {\bf 86}, 4783 (2001).

\bibitem{Song:2013qma} 
  H.~Song, S.~Bass and U.~W.~Heinz,
  Phys.\ Rev.\ C {\bf 89}, no. 3, 034919 (2014)
  doi:10.1103/PhysRevC.89.034919
  [arXiv:1311.0157 [nucl-th]].
  
\bibitem{Zhu:2015dfa} 
  X.~Zhu, F.~Meng, H.~Song and Y.~X.~Liu,
  Phys.\ Rev.\ C {\bf 91}, no. 3, 034904 (2015)
  doi:10.1103/PhysRevC.91.034904
  [arXiv:1501.03286 [nucl-th]].

\bibitem{Hirano:2005xf} 
  T.~Hirano, U.~W.~Heinz, D.~Kharzeev, R.~Lacey and Y.~Nara,
  Phys.\ Lett.\ B {\bf 636}, 299 (2006)
  doi:10.1016/j.physletb.2006.03.060
  [nucl-th/0511046].
	
\bibitem{Hirano:2007ei} 
  T.~Hirano, U.~W.~Heinz, D.~Kharzeev, R.~Lacey and Y.~Nara,
  Phys.\ Rev.\ C {\bf 77}, 044909 (2008)
  doi:10.1103/PhysRevC.77.044909
  [arXiv:0710.5795 [nucl-th]].
  
\bibitem{Hirano:2010jg} 
  T.~Hirano, P.~Huovinen and Y.~Nara,
  Phys.\ Rev.\ C {\bf 83}, 021902 (2011)
  doi:10.1103/PhysRevC.83.021902
  [arXiv:1010.6222 [nucl-th]].

\bibitem{Takeuchi:2015ana} 
  S.~Takeuchi, K.~Murase, T.~Hirano, P.~Huovinen and Y.~Nara,
  Phys.\ Rev.\ C {\bf 92}, no. 4, 044907 (2015)
  doi:10.1103/PhysRevC.92.044907
  [arXiv:1505.05961 [nucl-th]].

\bibitem{Nonaka:2007hy}
  C.~Nonaka and S.~A.~Bass,
  Phys.\ Rev.\ C {\bf 75}, 014902 (2007).

\bibitem{Petersen:2008in}
  H.~Petersen, J.~Steinheimer, G.~Burau, M.~Bleicher, and H.~St\"ocker
  Phys.\ Rev.\ C {\bf 78}, 044901 (2008).

\bibitem{Nonaka:2010hc}
  C.~Nonaka,
  AIP Conf.\ Proc.\ {\bf 1235} 165-171 (2010).

\bibitem{Petersen:2010tr}
  H.~Petersen, G.-Y.~Qin, S.~A.~Bass, and B.~M\"uller,
  Phys.\ Rev.\ C {\bf 82}, 041901 (2010).

\bibitem{Bartels:2002msat}
  J.~Bartels, K.~Golec-Biernat and H.~Kowalski,
  Phys.\ Rev.\ D {\bf 66}, 014001 (2002).

\bibitem{Kowalski:2003ipsat}
  H.~Kowalski and D.~Teaney,
  Phys.\ Rev.\ D {\bf 68}, 114005 (2003).

\bibitem{Kowalski:2006dip}
  H.~Kowalski, L.~Motyka and G.~Watt,
  Phys.\ Rev.\ D {\bf 74}, 074016 (2006).
  
\bibitem{Krasnitz:1999wc} 
  A.~Krasnitz and R.~Venugopalan,
  Phys.\ Rev.\ Lett.\  {\bf 84}, 4309 (2000)
  doi:10.1103/PhysRevLett.84.4309
  [hep-ph/9909203].

\bibitem{Krasnitz:2000gz} 
  A.~Krasnitz and R.~Venugopalan,
  Phys.\ Rev.\ Lett.\  {\bf 86}, 1717 (2001)
  doi:10.1103/PhysRevLett.86.1717
  [hep-ph/0007108].

\bibitem{Krasnitz:2002mn} 
  A.~Krasnitz, Y.~Nara and R.~Venugopalan,
  Nucl.\ Phys.\ A {\bf 717}, 268 (2003)
  doi:10.1016/S0375-9474(03)00636-5
  [hep-ph/0209269].

\bibitem{Schenke:2012fw} 
  B.~Schenke, P.~Tribedy and R.~Venugopalan,
  Phys.\ Rev.\ C {\bf 86}, 034908 (2012).


\bibitem{Huovinen:2009yb} 
  P.~Huovinen and P.~Petreczky,
  Nucl.\ Phys.\ A {\bf 837}, 26 (2010).

\bibitem{Denicol:2012cn} G.~S.~Denicol, H.~Niemi, E.~Molnar and
D.~H.~Rischke, 
Phys.\ Rev.\ D \textbf{85}, 114047 (2012). 

\bibitem{Cooper:1974mv} 
  F.~Cooper and G.~Frye,
  Phys.\ Rev.\ D {\bf 10}, 186 (1974).


\bibitem{Dusling:2010rel} 
  K.~Dusling, G.~D.~Moore and D.~Teaney
  Phys.\ Rev.\ C {\bf 81}, 034907 (2010).
  %
	
\bibitem{Paquet:2015lta}
  J.~F.~Paquet, C.~Shen, G.~S.~Denicol, M.~Luzum, B.~Schenke,
  S.~Jeon and C.~Gale,
  Phys.\ Rev.\ C {\bf 93}, no. 4, 044906 (2016)
  doi:10.1103/PhysRevC.93.044906
  [arXiv:1509.06738 [hep-ph]].


\bibitem{ALICE:2011ab} 
  K.~Aamodt {\it et al.}  [ALICE collaboration],
  Phys.\ Rev.\ Lett.\  {\bf 107}, 032301 (2011).

\bibitem{Bilandzic:2010jr} 
  A.~Bilandzic, R.~Snellings and S.~Voloshin,
  Phys.\ Rev.\ C {\bf 83}, 044913 (2011).

\bibitem{Pan:2014caa} 
  Y.~Pan and S.~Pratt,
  Phys.\ Rev.\ C {\bf 89}, no. 4, 044911 (2014).
  doi:10.1103/PhysRevC.89.044911

\bibitem{Abelev:2013vea} 
  B.~Abelev {\it et al.}  [ALICE Collaboration],
  Phys.\ Rev.\ C {\bf 88}, no. 4, 044910 (2013).

\bibitem{Abelev:2008ab} 
  B.~I.~Abelev {\it et al.} [STAR Collaboration],
  Phys.\ Rev.\ C {\bf 79}, 034909 (2009)

\bibitem{Adams:2004bi} 
  J.~Adams {\it et al.} [STAR Collaboration],
  Phys.\ Rev.\ C {\bf 72}, 014904 (2005)
  doi:10.1103/PhysRevC.72.014904
  [nucl-ex/0409033].
	
\bibitem{Adamczyk:2013waa} 
  L.~Adamczyk {\it et al.} [STAR Collaboration],
  Phys.\ Rev.\ C {\bf 88}, no. 1, 014904 (2013)
  doi:10.1103/PhysRevC.88.014904
  [arXiv:1301.2187 [nucl-ex]].

\bibitem{Abelev:2008ab} 
  B.~I.~Abelev {\it et al.} [STAR Collaboration],
  Phys.\ Rev.\ C {\bf 79}, 034909 (2009)
  doi:10.1103/PhysRevC.79.034909
  [arXiv:0808.2041 [nucl-ex]].

\bibitem{Bleicher:1999pu} 
  M.~J.~Bleicher, S.~A.~Bass, L.~V.~Bravina, W.~Greiner, S.~Soff, H.~Stoecker,
N.~Xu and E.~E.~Zabrodin,
  Phys.\ Rev.\ C {\bf 62}, 024904 (2000)
  [hep-ph/9911420].

\bibitem{Bratkovskaya:2000qy} 
  E.~L.~Bratkovskaya, W.~Cassing, C.~Greiner, M.~Effenberger, U.~Mosel and
A.~Sibirtsev,
  Nucl.\ Phys.\ A {\bf 675}, 661 (2000)
  [nucl-th/0001008].

\bibitem{ALICE:2014v2id}
  B.~Abelev {\it et al.}  [ALICE collaboration],
  JHEP {\bf 06}, 190 (2015).

\bibitem{PHENIX:2004idsp}
  S.~S.~Adler {\it et al.} [PHENIX collaboration],
  Phys.\ Rev.\ C {\bf 69}, 034909 (2004).

\bibitem{STAR:2005vn}
  J.~Adams {\it et al.} [STAR collaboration],
  Phys.\ Rev.\ C {\bf 72}, 014904 (2005).

\bibitem{ALICE:2014mstr}
  [ALICE collaboration],
  Phys. Lett. B {\bf 728} 216 (2014).

\bibitem{Abelev:2013Ks0L}
  B.~Abelev {\it et al.}  [ALICE collaboration],
  Phys.\ Rev.\ Lett.\ {\bf 111}, 222301 (2013).

\bibitem{vanHecke:1999jh} 
  H.~van Hecke, H.~Sorge and N.~Xu,
  Nucl.\ Phys.\ A {\bf 661}, 493 (1999).

\bibitem{Arbex:2001vx} 
  N.~Arbex, F.~Grassi, Y.~Hama and O.~Socolowski,
  Phys.\ Rev.\ C {\bf 64}, 064906 (2001).

\bibitem{Chatterjee:2013yga} 
  S.~Chatterjee, R.~M.~Godbole and S.~Gupta,
  Phys.\ Lett.\ B {\bf 727}, 554 (2013)
  [arXiv:1306.2006 [nucl-th]].

\bibitem{Bugaev:2016voh} 
  K.~A.~Bugaev {\it et al.},
  Ukr.\ J.\ Phys.\  {\bf 61}, 659 (2016)
  [arXiv:1610.03269 [nucl-th]].

\bibitem{Huovinen:2008ch}
  P.~Huovinen,
  Eur.\ Phys.\ J.\ A {\bf 37}, 121 (2008).

\bibitem{CMS:2013v2}
  S.~Chatrchyan {\it et al.}  [CMS collaboration],
  Phys.\ Rev.\ C {\bf 87}, 014902 (2013).

\bibitem{CMS:2014vn}
  S.~Chatrchyan {\it et al.}  [CMS collaboration],
  Phys.\ Rev.\ C {\bf 89}, 044906 (2014).

\bibitem{Adam:2016nfo} 
  J.~Adam {\it et al.} [ALICE Collaboration],
  JHEP {\bf 1609}, 164 (2016)
  doi:10.1007/JHEP09(2016)164
  [arXiv:1606.06057 [nucl-ex]].

\bibitem{PHENIX:2011vnch}
  A.~Adare {\it et al.} [PHENIX collaboration],
  Phys.\ Rev.\ Lett. {\bf 107}, 252301 (2011).

\bibitem{STAR:2008vn}
  B.~I.~Abelev {\it et al.} [STAR collaboration],
  Phys.\ Rev.\ C {\bf 77}, 054901 (2008).

\bibitem{PHENIX:2004idsp}
  S.~S.~Adler {\it et al.} [PHENIX collaboration],
  Phys.\ Rev.\ C {\bf 69}, 034909 (2004).

\bibitem{STAR:2007st}
  J.~Adams {\it et al.} [STAR collaboration],
  Phys.\ Rev.\ Lett. {\bf 98}, 62301 (2007).

\bibitem{STAR:2016vs}
  L.~Adamczyk {\it et al.} [STAR collaboration],
  Phys.\ Rev.\ Lett. {\bf 116}, 62301 (2016).

\bibitem{Shen:2012us} 
  C.~Shen and U.~Heinz,
  Nucl.\ Phys.\ A {\bf 904-905}, 361c (2013)
  doi:10.1016/j.nuclphysa.2013.02.024
  [arXiv:1210.2074 [nucl-th]].
	
\bibitem{Molnar:2014fva} 
  D.~Molnar and Z.~Wolff,
  Phys.\ Rev.\ C {\bf 95}, no. 2, 024903 (2017)
  doi:10.1103/PhysRevC.95.024903
  [arXiv:1404.7850 [nucl-th]].

\end{thebibliography}
\end{document}